\documentclass[prb,showpacs,superscriptaddress,groupedaddress,twocolumn]{revtex4-1}
\usepackage{amssymb}
\usepackage{amsmath}
\usepackage{color}
\usepackage{epstopdf}
\usepackage{graphicx}
\usepackage{soul}

\def\nn{\nonumber}

\def\mb{\mathbf}
\def\be{\begin{equation}}
\def\ee{\end{equation}}
\def\ba{\begin{eqnarray}}
\def\ea{\end{eqnarray}}

\newcommand{\de}[1]{\textcolor{green}{#1}}
\begin{document}
\title{Anomalous diamagnetic response in multi-band superconductors with time-reversal broken symmetry.}

\author{Yuriy Yerin}
\affiliation{B.\ Verkin Institute for Low Temperature Physics and Engineering,
National Academy of Sciences of Ukraine, 47 Lenin Ave., 61103, Kharkiv,
Ukraine}
\affiliation{Institute for Physics of Microstructures,
Russian Academy of Sciences,
603590 Nizhny Novgorod, GSP 105,
Russia}
\author{ Alexander Omelyanchouk}
\affiliation{B.\ Verkin Institute for Low Temperature Physics and Engineering,
National Academy of Sciences of Ukraine, 47 Lenin Ave., 61103, Kharkiv,
Ukraine}
\author{ Stefan-Ludwig Drechsler}
\affiliation{Institute for Theoretical Solid State Physics, Leibniz-Institut f\"ur Festk\"orper-
und Werkstoffforschung IFW-Dresden, D-01169 Dresden, Helmholtzstra\"sse 20,
Germany}
\author{Dmitri V.\ Efremov}
\affiliation{Institute for Theoretical Solid State Physics, Leibniz-Institut f\"ur Festk\"orper-
und Werkstoffforschung IFW-Dresden, D-01169 Dresden, Helmholtzstra\"sse 20,
Germany}
\author{ Jeroen van den Brink}
\affiliation{Institute for Theoretical Solid State Physics, Leibniz-Institut f\"ur Festk\"orper-
und Werkstoffforschung IFW-Dresden, D-01169 Dresden, Helmholtzstra\"sse 20,
Germany}

\begin{abstract}
Within a Ginzburg-Landau formalism we establish analytically the necessary and sufficient conditions to realize a doubly degenerate superconducting ground state with broken time-reversal symmetry (BTRS) in a multi-band superconductor.
%
%
Using these results we analyze the ground state of a three band superconductor in the cylindrical geometry in an external magnetic field.
%
%
We show that depending on the interband coupling constants,  a magnetic flux can induce current density jumps in such superconducting geometries that are related to adiabatic or non-adiabatic
transitions from BTRS to time-reversal symmetric states and vice versa.
This unusual current induced magnetic flux response can in principle be used experimentally to detect superconducting BTRS ground states as well as corresponding metastable excited states.
\end{abstract}

\date{\today}

\maketitle

\section{Introduction}
The phenomenon of  superconductivity based  is characterized by the spontaneous breaking of a gauge symmetry.
But in some cases simultaneously  time-reversal symmetry (TRS) can be broken as well. Because of their unusual properties such superconductors (SC) with
broken time-reversal symmetry (BTRS)  are attracting a lot of attention. For instance recently the formation of new collective modes \cite{Marciani2013,Carlstrom2011}
(similarly to the occurrence of Leggett modes in two-band SC) and new topological excitations in the form of phase kinks, domains and vortices that carry fractional magnetic flux values
\cite{Bojesen2013,Shi-Zheng2012,Yanagisawa2012,Lin2014,Tanaka2015,Chubukov2016,Huang2016,Garaud2016,Koyama2016,Stanev2015}
have been discussed.

So far, BTRS superconductivity has been clearly detected in few cases, only. The most frequently cited example is  Sr$_2$RuO$_4$ in which the order parameter was identified to be
triplet chiral ($\Delta \propto p_x+i p_y$) \cite{Mackenzie2003,luketime1998}. Furthermore, evidence for  BTRS superconductivity was reported for the low-$T$ phase of Th-doped UBe$_{13}$ \cite{Heffneret},  UPt$_3$ \cite{Lukeet}, and in  SrPtAs based on muon measurements \cite{Biswas}. Theoretically superconductivity with BTRS is also been proposed for other compounds such as cuprate SC at low-temperature \cite{Laughlin1998,Krishana1997}, transition metal dichalcogenides \cite{Efremov2014,Thomale2014}, Na$_x$CoO$_2 \cdot y$H$_2$O \cite{Thomale2013}, strongly doped graphene \cite{Nandkishore2012}, and in some recently discovered Fe-based superconductors (FeSC).

The FeSC are of particular interests as BTRS superconductivity is anticipated for several dopings as a result of the multiband electron structure and strong repulsive interband couplings. Typically the Fermi-surface of  the non-SC parent compound consists of two or three hole-like pockets at $\Gamma = (0,0)$ point and two electron-like pockets around $M=(\pi,\pi)$ point. The resulting nesting at the vector $Q = (\pi, \pi)$ connecting the $\Gamma$ and M points drives the system to a spin-density wave (SDW) state. With doping the SDW state melts, giving space for superconductivity. The natural symmetry of the order parameter  in such a situation is given by the so called $s_\pm$ one, which causes a gap function with opposite signs at the electron and the hole pockets, respectively. In Ba$_{1-x}$K$_x$Fe$_2$As$_2$ the hole doping by K-substitution leads at x close to 1 to the vanishing of the electrons pockets and a
change of the symmetry of the superconducting order parameter to nodal $d$-wave  symmetry for pure KFe$_2$As$_2$ \cite{Tafti2013,Terashima2013,Abdel-Hafiez2013,Grinenko2014}.
A BTRS  state  was proposed for two dopings: for $x\sim 0.7$, when the electron pockets vanish,  and for $x\sim 1$, when a transition from $s$ to $d$-wave superconductivity is expected. The proposed intermediate pairing symmetries  are  $s+is$ and  $s+id$ correspondingly \cite{Wei-Cheng2009,Stanev2014}.

However, as stressed above, still there are only few compounds where superconductivity with BTRS was unambiguously observed. One of the reasons actually is the lack of simple experimental tools to identify it. The most common techniques to establish BTRS in superconductors are $\mu$SR and NMR both suffer from restrictions related to the the presence of impurities and other defects even in high-quality single crystals. In addition NMR requires considerable magnetic fields which may themselves significantly affect the superconducting GS, especially for low-temperature SC by paramagnetic pair-breaking effects and field induced coexisting magnetic phases.

Here we first study in Sect. II the nature of possible SC ground states in zero magnetic field and absence of currents. Within a Ginzburg-Landau approach we clarify the conditions for BTRS and explicitly show the two-fold degeneracy of the corresponding ground states. We adopt the simplest  model when the BTRS superconductivity, namely a three-band SC, is described approximately within a Ginzburg-Landau approach \cite{Xiao2012,Dias2011,Wilson2013,Tanaka2013,Stanev2010,Yerin2013}. For two-band model BTRS superconductivity is possible only in special cases like dirty materials in the vicinity  of the $s_{\pm} \to s_{++}$ transition \cite{Stanev2014}.

In Sect. III we investigate the magnetic field response of superconducting cylinders with a BTRS order parameter and introduce it as a new tool to identify superconductivity with BTRS. In the three-band framework we investigate the homogeneous current states in such a mesoscopically one-dimensional system and show that the diamagnetic i.e.\ an orbital dominated response depends directly on the nature of the underlying order parameter. Experimental verification of characteristics that we predicted here could be used to identify multiband BTRS-superconductivity.


\section{Ginzburg-Landau approach to  superconductivity with BTRS}
To describe the multiband superconductors we employ a general Ginzburg-Landau (GL) functional, which has been used previously for  particular cases (e.g.\ for two-bands or three equivalent bands, \cite{Zhitomirsky2004,Moor2013} see the reviews Refs. [\onlinecite{Lin2014}], [\onlinecite{Tanaka2015}] and references therein), only.
We will provide a rather general solution for three {\it non-equivalent}  bands with repulsive interband interactions being the most relevant case for BTRS-physics in these systems. For the sake of simplicity we address only homogeneous states and isotropic order parameters. However, inhomogeneous states containing different topological defects and explicit account of spin states can be treated straightforwardly within the same formalism \cite{remark2}.
%
The Ginzburg-Landau (GL) Gibbs energy density and the current density for three-band superconductors  can be written in the following form
\begin{eqnarray}
\Delta G &= \sum_{i=1}^3 \int \left\{ \frac{1}{4m_i}\left|\left(-i\hbar \nabla
-\frac{2e}{c} \mb{A}\right) \psi_i \right|^2+ a_i |\psi_i|^2 \right. \nonumber  \\
& \left. + \frac{b_i}{2} |\psi_i|^4
- F_{int} \right\}  + \frac{1}{8\pi} \int (rot A - H)^2
\label{eq.GL}
\end{eqnarray}
and
\be
\mb{j} =   -  \sum_i \frac{ i e \hbar}{ m_i} \left(\psi_i^* \nabla \psi_i -
 \psi_i \nabla \psi_i^*\right) - \frac{4e^2}{ c} \mb{A}\sum_i \frac{|\psi_i|^2}{m_i}.
\label{eq.current}
\ee
Here and below we consider the temperature regime below $T_c$ for the GL approach is valid, i.e. we ignore the region of strong fluctuations in the very vicinity of $T_c$ or the case of very low temperatures. In the first integral in  Eq.(\ref{eq.GL}) the  integration is performed  over the superconductive region whereas in the second integral over the cylinder volume. The  term  $F_{int}$  describes the phase sensitive Josephson-like interband coupling:
\ba
  F_{int} &=& \gamma_{12} \psi_1^\star \psi_2+  \gamma_{23} \psi_2^\star \psi_3
    +\gamma_{31}  \psi_1^\star \psi_3  + c.c.\  \quad  \nn \\
  &=& 2\gamma_{12} |\psi_1| |\psi_2| \cos( \phi_1 - \phi_2 ) \nn \\ &+&   2\gamma_{23} |\psi_2| |\psi_3| \cos( \phi_2 - \phi_3 )   \\ &+& 2 \gamma_{31} | \psi_1| |\psi_3| \cos( \phi_1 - \phi_3 ) \nn
    \label{eq.fint}
\ea
Here the order parameter is in general complex, i.e.\
$\psi_i = |\psi_i| e^{i \phi_i} $. In contrast,
the interaction coefficients $\gamma_{ij}$  are real and
can be positive or negative.
It was shown for two-band superconductors, that the sign of $\gamma_{12}$
fully determines the symmetry of the order parameter in the clean case.
A repulsive interband
 interaction  constant $\gamma_{12}<0$  leads to unconventional
 symmetry and a ground state
 with $\pi$-phase difference between the two bands (denoted as
 $s_{\pm}$ -symmetry),
    while attractive interband interactions
    $\gamma_{12}>0$  stabilize a
ground state with a zero-phase difference between the their gap
functions
(denoted as $s_{++}$-symmetry).
We keep the same sign-convention in
the case of
three-band SCs considered here.
\begin{figure}
\includegraphics[width=0.49\columnwidth]{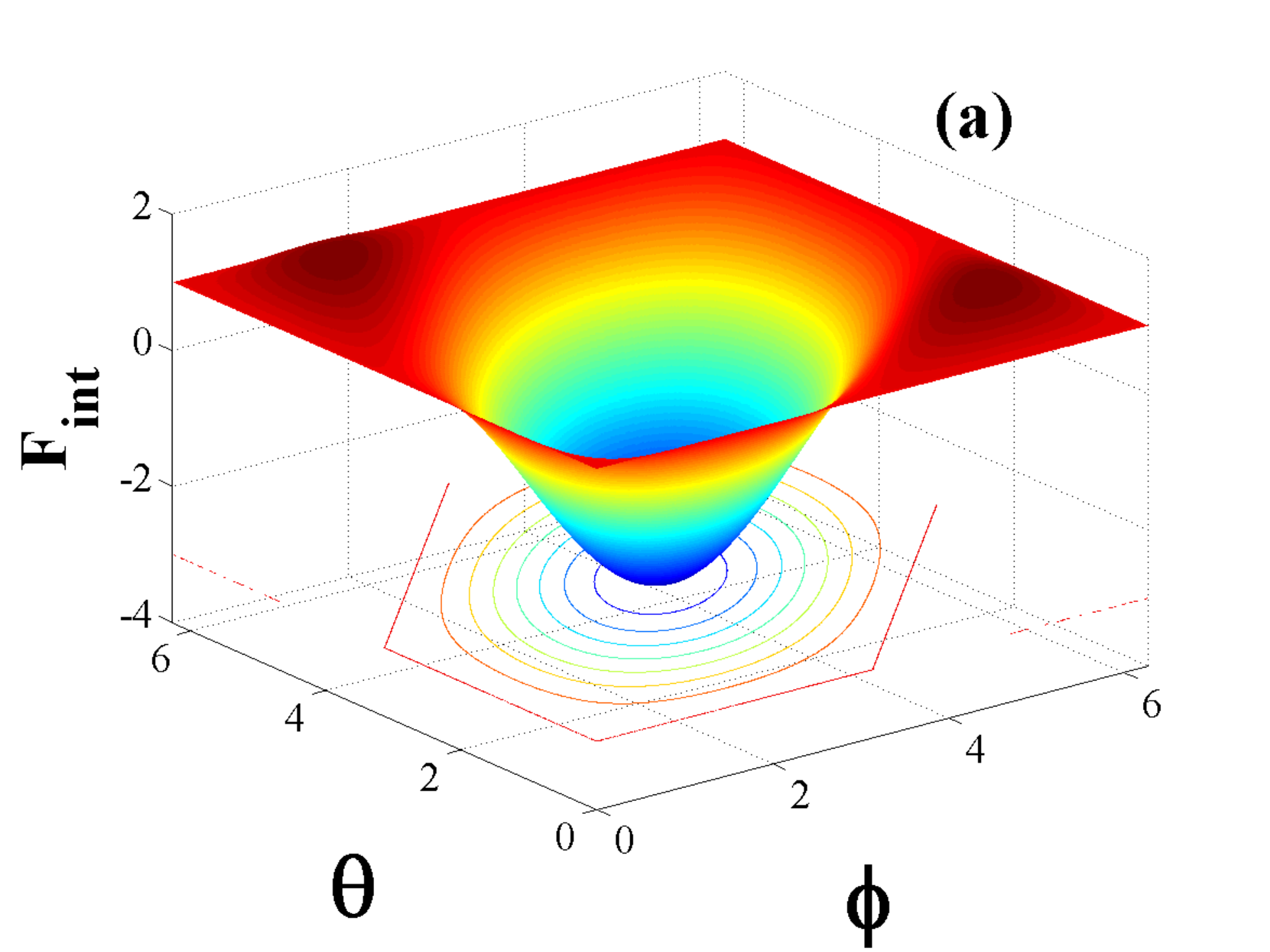}
\includegraphics[width=0.49\columnwidth]{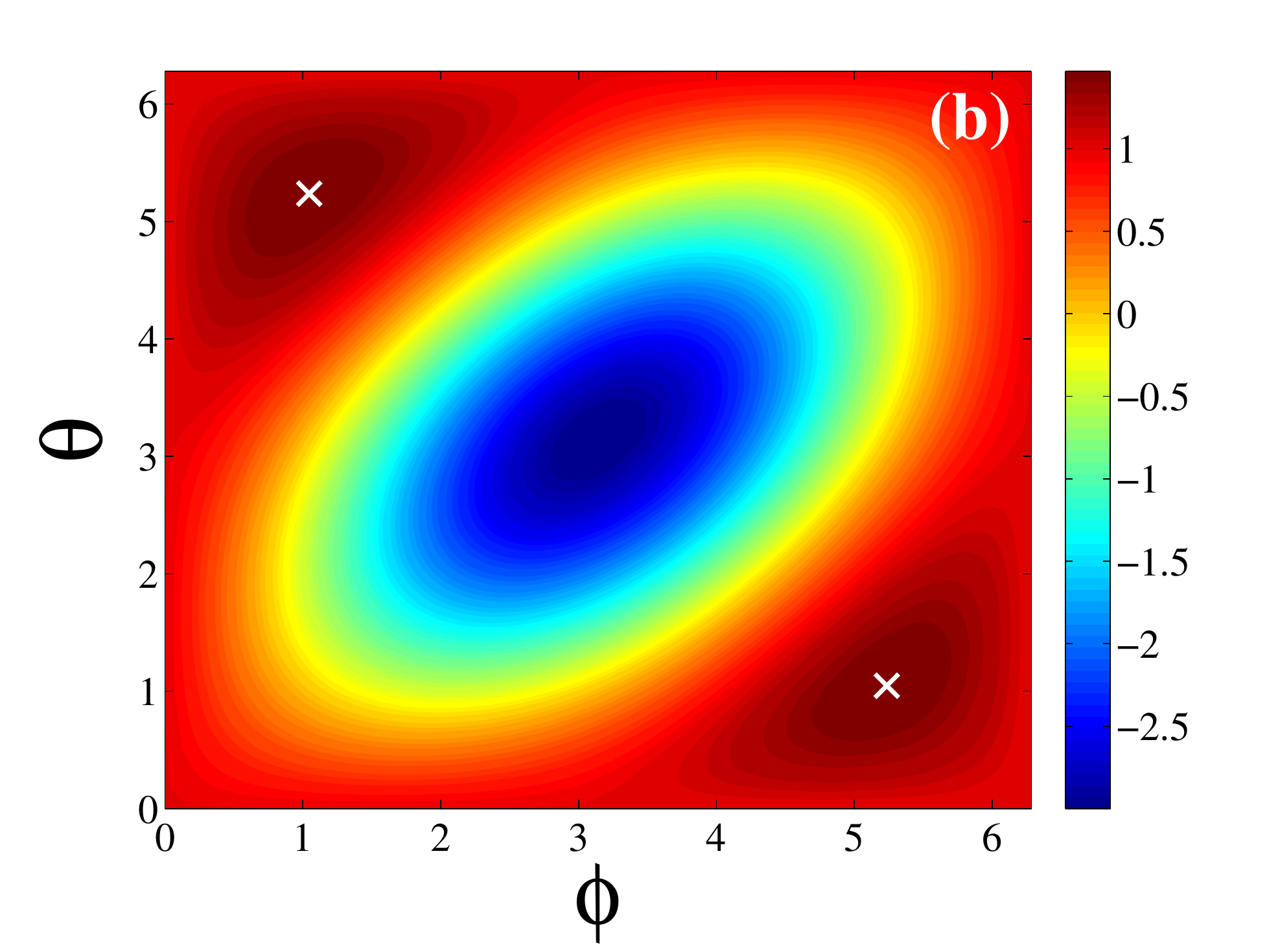}
\caption{The surface (a) and the contour plot (b) of
the dependence of the interband interaction term $F_{int}$ of the GL free
energy functional as a function of the phase differences for the fixed set
$G_1=1$, $G_2= -1$, $G_3=1$.
 Black dashed square
limits intervals of consideration of phase differences.
Crosses indicate global maximum values of $F_{int}$ within
these intervals. The parameters are $\phi =\phi_1 - \phi_2 $ and $\theta = \phi_1 - \phi_3  $.
Further details are provided in Supplement, Figs.\ S4.
}
\label{fig1}
\end{figure}

As a first step we examine the ground state in the absence of external
magnetic fields. Then eq.\ (\ref{eq.fint}) can be rewritten as
$F_{int} = \operatorname{sgn}(\gamma_{12}\gamma_{23}\gamma_{31} )(F_x^2 +
F_y^2 - \gamma_1^2 |\psi_1|^2 - \gamma_2^2 |\psi_2|^2- \gamma_3^2 |\psi_3|^2) $, where
$\gamma_{1} =  \sqrt{| \gamma_{12}\gamma_{23}\gamma_{31} |}/ \gamma_{23}$
and the other $\gamma_i$ are obtained by a cycle permutation. The introduced two functions
$F_x = \gamma_1 |\psi_1|\cos \phi_1 + \gamma_2 |\psi_2|\cos \phi_2 + \gamma_3 |\psi_3|\cos \phi_3   $
and $F_y = \gamma_1 |\psi_1|\sin \phi_1 + \gamma_2 |\psi_2|\sin \phi_2 + \gamma_3 |\psi_3|\sin \phi_3 $,
do absorb the complete phase shift dependencies.

Now the minimization of the GL-functional  with respect to the phases is
reduced to the maximization/minimization of $F_x^2+F_y^2$ depending on the
sign of $(\gamma_{12}\gamma_{23}\gamma_{31} )$. The geometric meaning of
$F_x^2+F_y^2$ is the absolute value of a sum of three vectors  in a
2D space $\mathbf{a}_1=\gamma_1 |\psi_1| (\cos \phi_1, \sin \phi_1)$,
$\mathbf{a}_2=\gamma_2 |\psi_2| (\cos \phi_2, \sin \phi_2)$ and
$\mathbf{a}_3=\gamma_3 |\psi_3| (\cos \phi_3, \sin \phi_3)$: $F_x^2 + F_y^2 = (\mathbf{a}_1+\mathbf{a}_2+\mathbf{a}_3)^2 $. One can immediately note that BTRS state corresponds to noncollinear  vectors $\mathbf{a}_i$, while  TRS to collinear ones.  For $(\gamma_{12}\gamma_{23}\gamma_{31} )>0$ the minimum of GL corresponds to the maximum of the $F_x^2 +F_y^2$, which is reached when the vectors $\mathbf{a}_i$ are collinear. It corresponds to the TRS phase. For $(\gamma_{12}\gamma_{23}\gamma_{31} )<0$   the minimum of GL corresponds to the minimum of the $F_x^2 +F_y^2$. The minimum $F_x^2+F_y^2 = 0$  can be reached for noncollinear vectors, satisfying the triangle rule. With this the BTRS GS is realized.
If the length of vectors $\mathbf{a}_i$ does not satisfy the triangle rule the minimum is reached for collinear vectors $\mathbf{a}_i$, or TRS state.
For the BTRS state the free energy $\Delta F_{BTRS}$ can be rewritten as:
\be
\Delta F_{BTRS} = \sum_i [(a_i - \gamma_i^2) |\psi_i|^2 + \frac{b_i}{2}|\psi_i|^4 ].
\label{eq.GL.BTRS}
\ee
Its minimum corresponds to the order parameter:
\be |\psi_i|^2 = (-a_i+\gamma_i^2)/b_i  \label{eq.psi.BTRS} \ee
and
\ba
\cos \theta=\cos (\phi_2 -\phi_1) = \frac{\gamma_3^2 |\psi_3|^2-\gamma_1^2 |\psi_1|^2-\gamma_2^2
|\psi_2|^2}{2 \gamma_{12} |\psi_1| |\psi_2|}, \label{eq.cos1} \nn \\
\cos \phi=\cos (\phi_3 -\phi_1) = \frac{\gamma_2^2 |\psi_2|^2-\gamma_1^2
|\psi_1|^2-\gamma_3^2 |\psi_3|^2}{2 \gamma_{13} |\psi_1| |\psi_3|}. \label{eq.cos2} \nn
\ea

\begin{figure}
\includegraphics[width=0.49\columnwidth]{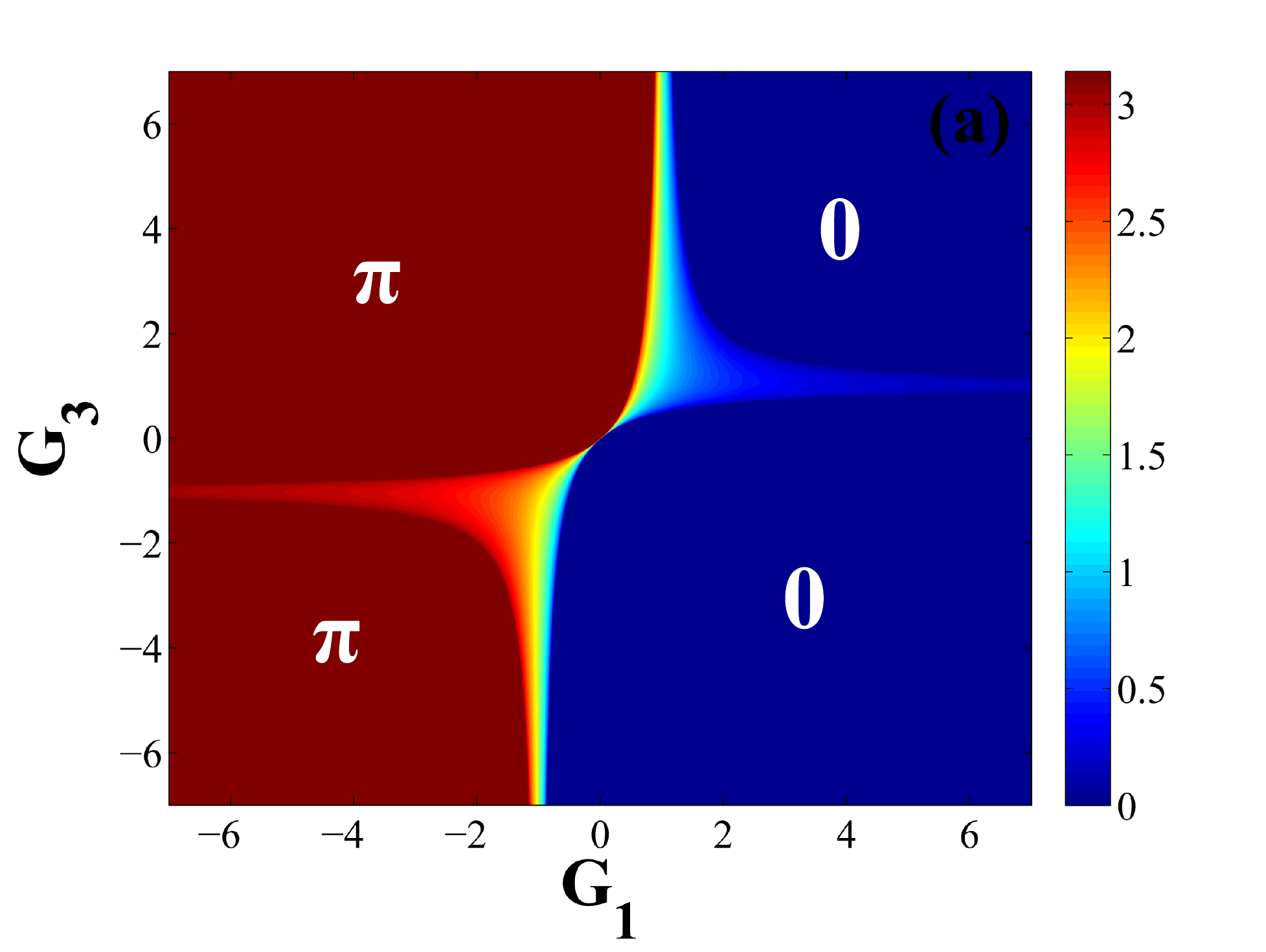}
\includegraphics[width=0.49\columnwidth]{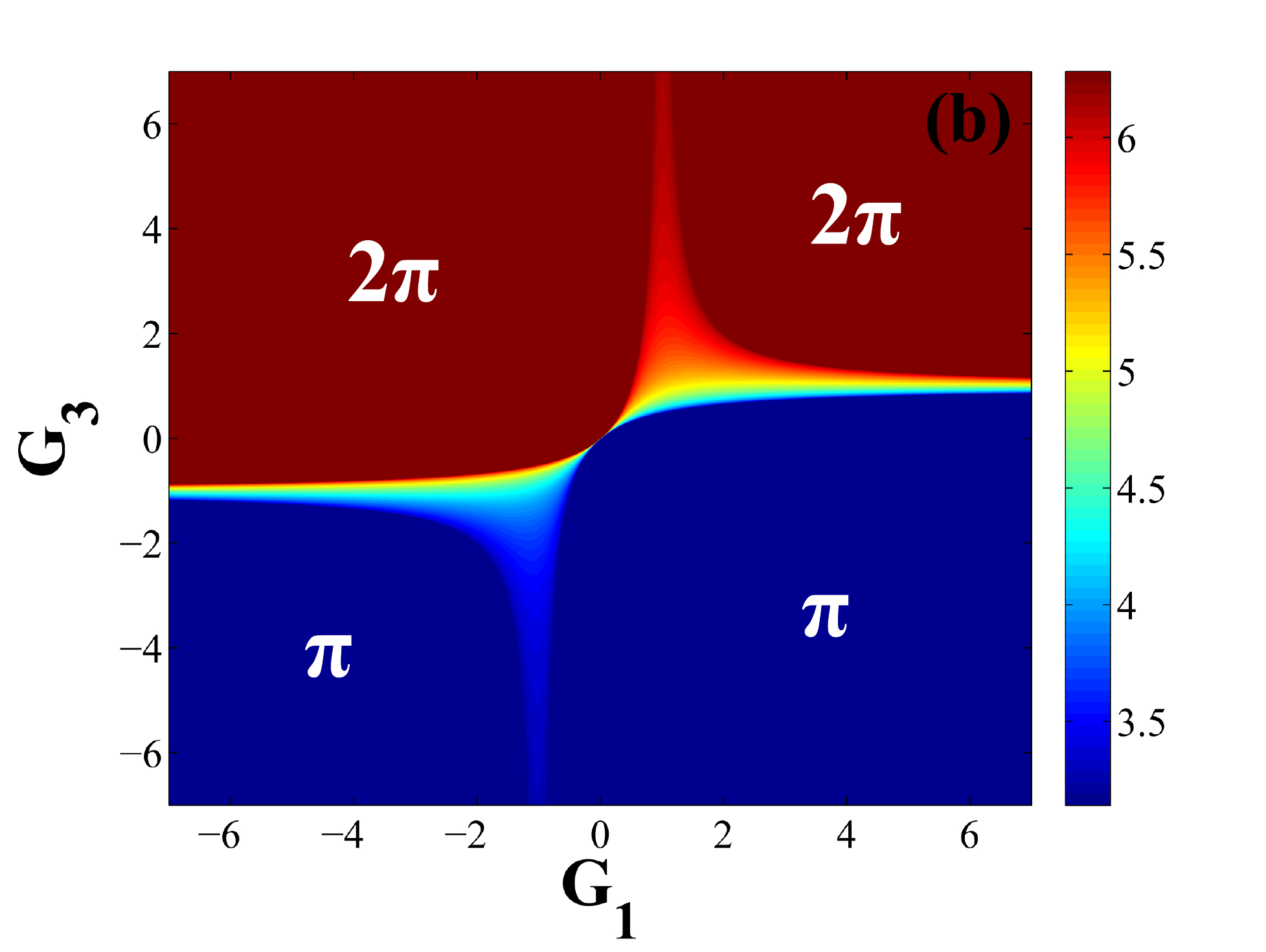}
\caption{Regions of existence of a BTRS ground state
in terms of the two phase variables
for $\phi$  (upper panel) and $\theta$  (lower panel)
 of a three-band SC .
The BTRS ground states are realized in the narrow
"tilted x-like" regions with inhomogeneous
colors ($\theta,\phi \neq 0,\pi$), only, distinct from the large homogeneous regions corresponding to TRS ground states. The parameters are $\phi =\phi_1 - \phi_2 $ and $\theta = \phi_1 - \phi_3  $
For more details see Supplement, Figs.\ S2 and S3.}
 \label{fig3}
\end{figure}
The dependence of the interband interaction term $F_{int}$ of the GL free
energy functional as a function of the phase differences for a fixed set
$G_i $ is given in Fig. \ref{fig1}.
Here we have introduced convenient new variables:
$$
G_1 = \frac{\gamma_{12}}{|\gamma_{23}|} \frac{|\psi_1|}{|\psi_2|} \mbox{,~~}
G_2 = \textrm{sign}(\gamma_{23}) \mbox{~~and~~} G_3 = \frac{\gamma_{13}}{|\gamma_{23}|} \frac{|\psi_3|}{|\psi_2|}.
$$
A BTRS state exists for zero external magnetic field
only within a relative small volume in the six-dimensional
parameter space ($| \psi_j | $,$\gamma_{ij}$). The corresponding projected
regions onto the planes $G_1$-$G_3$ are
shown in the form of  "tilted X-like" regions in Fig.~\ref{fig3}.
The richness of these and other figures shown in the Supplement is a consequence of the high-dimensionality of the parameter space that is generic for multi-band superconductors. We note that it resembles mathematically to some extent the richness of the 11-dimensional superstring theory manifested in the six-dimensional Calabi-Yau manifolds \cite{Hanson1994,Candelas1985}.

Noteworthy  we have found that even in the case of an odd number of repulsive interband interactions, the degeneracy of ground states can be removed and TRS state can be stable. This means that the presence of one or three repulsive interband interactions in three-band superconductors does {\it not} provide a necessary and sufficient condition for the occurrence of a BTRS GS, contrary to some statements found in the literature \cite{tanaka2010b,tanaka2010c,Dias2011}.
%
To get a deeper insight into the nature of the interband frustration responsible for the appearance of BTRS  in three-band SCs, a rigorous and straightforward mathematical approach
is necessary (for details see the calculations and results presented graphically in the Figs. S1 and S2 of the Supplement.)
Correspondingly, for higher \textit{n}-band frustrated superconductors one is confronted with  $n\left(n-1\right)/2-1$ mutual phase differences, which can be considered within
the proposed geometrical interpretation. However, in contrast to the three band case the bilinear interaction between the superconducting
band order parameters is not enough for unambiguous determination of the phase differences and higher order terms have to be considered.

\section{BTRS and TRS superconductors on a cylinder and in magnetic field}
\begin{figure}
\includegraphics[width=0.7\columnwidth]{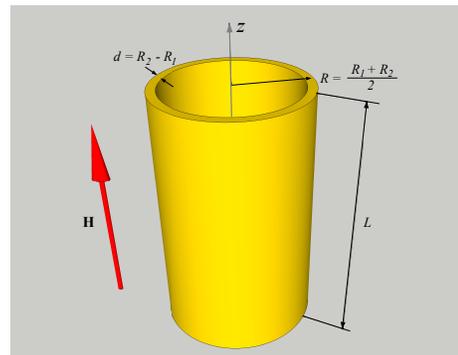}
\caption{(Color online)
Sketch of a tube made from a three-band superconductor with a BTRS ground state in a parallel external magnetic field.
}
\label{fig.tube}
\end{figure}

In order to distinguish readily a three-band SC
with a BTRS ground state from those
traditional SCs with a TRS
ground state, we propose to
apply a magnetic
flux to a (topologically) doubly-connected system.
In
particular, we consider a long and thin tube
approximated by two concentric cylinders
with the inner and outer radii $R_{1} $ and $R_{2} $,
 respectively, (see Fig.~\ref{fig.tube}).
Its symmetry axis is denoted as the \textit{z} axis in cylindrical
coordinates $\left(r,\vartheta ,z\right)$. An external constant magnetic
field $H$ is thought to be applied along the symmetry axis of such a
cylinder:$H=\left(0,0,H\right)$, where the vector-potential gauge is
chosen as  $A=\left(0,A_{\vartheta } \left(r\right),0\right)$,
$A_{\vartheta } \left(r\right)=\frac{Hr}{2} $.
We assume that the radius of the tube $R$ and the thickness $d$ satisfy
the following conditions: $R \gg \lambda, \xi$ and $d \ll \lambda, \xi$, where
$\lambda(T)$ and $\xi(T)$ are the London penetration depth and the coherence length respectively. The first
 condition  precludes  the formation of magnetic vortices or
any domains in the cylinder, while due the second one
the self-induced magnetic fields are small and can therefore be ignored
in our calculations. It means that we study only homogeneous
solutions $|\psi_i|= const$, while the phase depends on the polar angle $\vartheta$, only.
In the considered geometry the phase must fulfil the quantization condition   $\oint_\Gamma \nabla \phi_i = 2\pi n_i $,
where  the integral is taken over an arbitrary closed continuous contour
$\Gamma$  lying inside the cylinder and $n = 0, \pm1, \pm 2$  are the
phase winding (topological) numbers. Here we assume these winding numbers to be equal: $n_1 = n_2 = n_3$.

It is interesting to note that a similar experimental setup was proposed with the aim to detect a fractional flux plateau in the magnetization curve of a superconducting loop that is topologically the same as the one considered here\cite{Huang2015}. These authors investigate numerically metastable phase kinks protected by a large energy barrier within a GL functional adopting certain special parameter values.  The excited states with BTRS discussed in the following differ significantly from those solitonic states.

To illustrate the principle of  identifying BTRS for a given three-band SC, we consider a simple
case and assume firstly that the equilibrium values of the order parameters are given but
without adopting thereby the equality for the moduli of the interband interactions. Secondly,
the strengths of the interband interactions coincide but, for instance, at least one of these interactions is repulsive. Since we are interested in a
three-band superconductor with initial BTRS state we control the selection
of parameters of interband interactions numerically in order to avoid the
possible occurrence of non-frustrated ground states even for an odd number
of repulsive interband interaction (see Figs.\ S1-S8 in the Supplemental materials).
Then we can write the GL free-energy of the system in the momentum space
(see the Supplemental material)
\begin{widetext}
\begin{equation}
\frac{\Delta F}{\pi R^{2} L} =\sum _{i}\left( a_i |\psi_{i}| ^{2} +\frac{1}{2} b_i | \psi_{i}| ^{4}
+ \bar\kappa_i|\psi_{i}| ^{2} q^{2} \right) -2 {\gamma}_{12} |\psi_{1}| |\psi_{2}|
\cos \phi -2 {\gamma }_{13}|\psi_{1}|| \psi_{3} |\cos \theta
-2 {\gamma }_{23}|\psi_{2}|| \psi_{3} |\cos \left(\theta -\phi \right),
\label{eq.GLinMagneticField}
\end{equation}
\end{widetext}
and the current density is:
\be
j = \sum_i \bar\kappa_i |\psi_i|^2 q.
\label{eq.current_density}
\ee
Further we will use  $\kappa_i = \bar\kappa_i/\bar\kappa_1$. The
superfluid momentum
$q$ depends on the winding number $n$ and the magnetic flux $\Phi$ as
$q = \frac{1}{R} \left(n - \Phi/\Phi_0 \right)$  with
$\Phi_0 = \pi \hbar c/e$  being the flux quantum.
One sees
that Eq. (\ref{eq.GLinMagneticField}) can be obtained from the corresponding equation in zero magnetic field by substituting $a_i \to a_i + \kappa_i q^2$, i.e. an increase of $q$ acts in the same way as an increase of temperature.  With this remark we can apply the considered above results for  zero magnetic field.

\begin{figure}
\includegraphics[width=0.49\columnwidth]{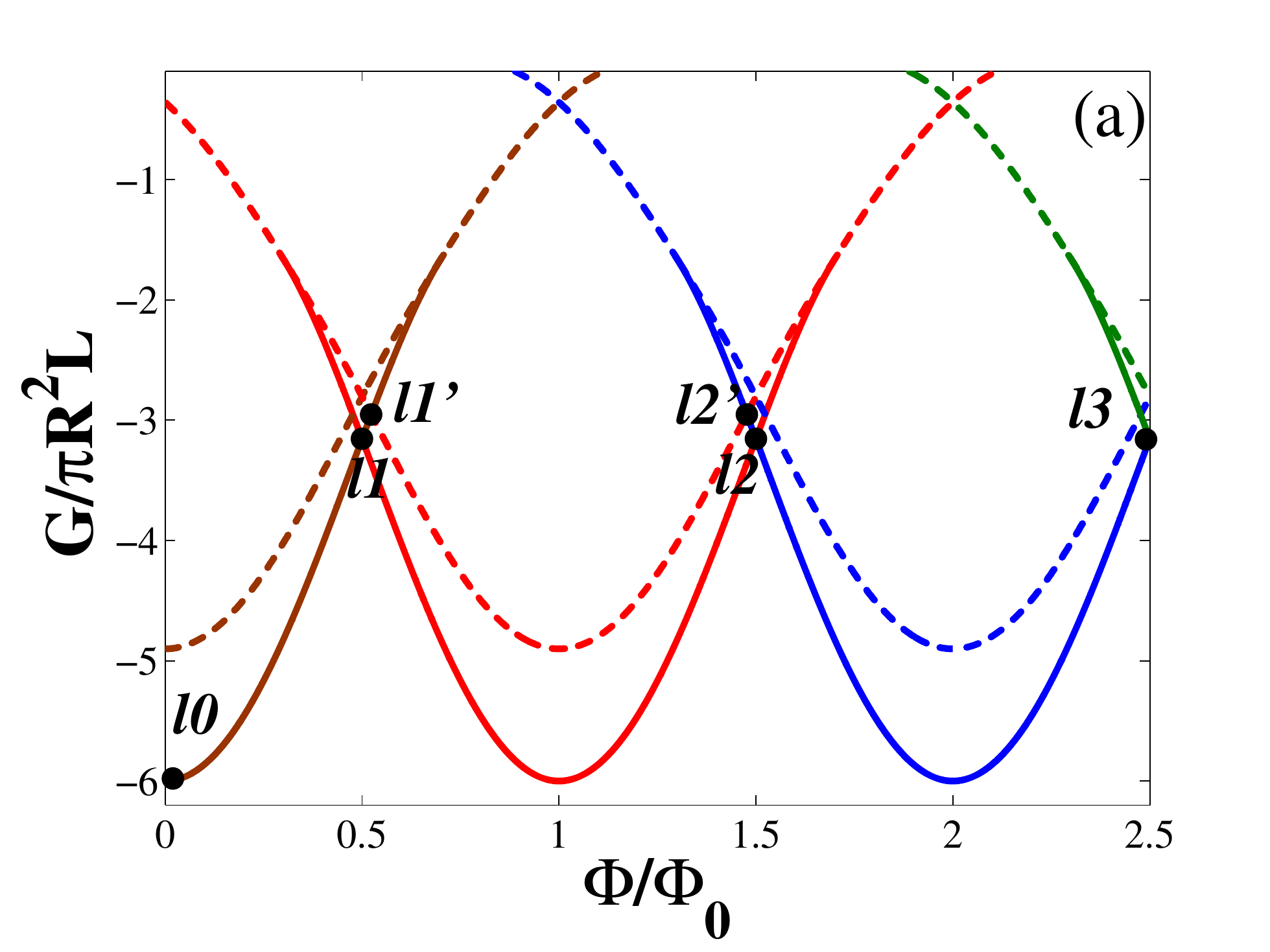}
\includegraphics[width=0.49\columnwidth]{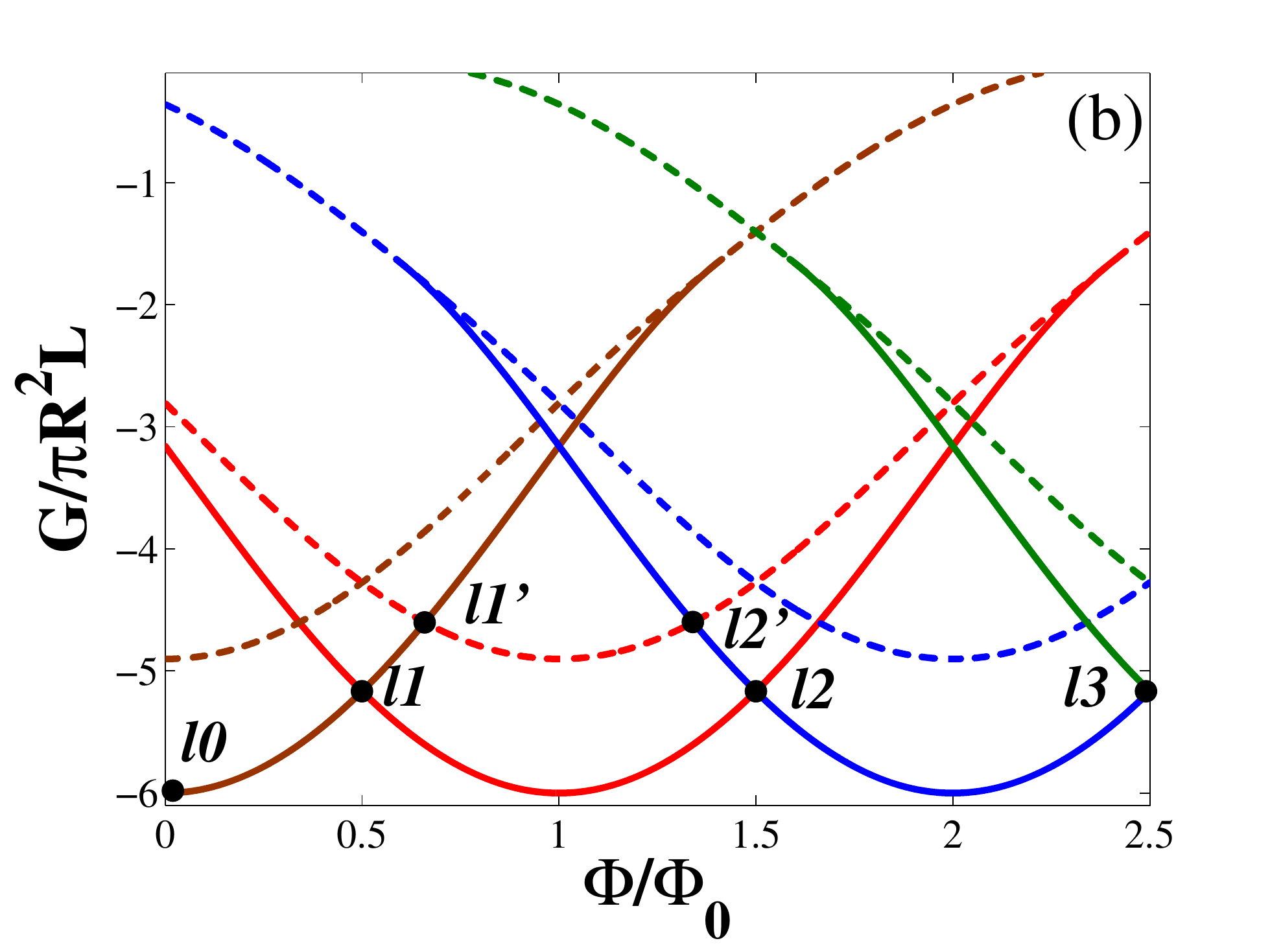}
\caption{Dependencies of the GL free energy  on the applied magnetic
flux for a three-band superconductor with
   $\tilde{\gamma }=1$  for  attractive interband interactions between
   the first and the second, the first and the third
 bands and a repulsive one between the second and the third bands,
 and for
 $\kappa_2 =4$  and $\kappa_3 =2$  (a) and $\kappa_2 =0.25$  and  $\kappa_3 =0.5$
 (b). Solid and dashed lines:
  GL free energies of a three-band superconductor with BTRS and without one,
   respectively, for different winding
 numbers $n=0$  (brown), $n=1$   (red),   $n=2$ (blue), $n=3$   (green) etc.
  The filled circle markers with captions
 determine possible ways of  evolution for the three-band superconducting
 long tube setup shown in Fig. 2
 in an external magnetic field
 (for explanations, see text).}
\label{fig.5}
\end{figure}

\begin{figure}
\includegraphics[width=0.7\columnwidth]{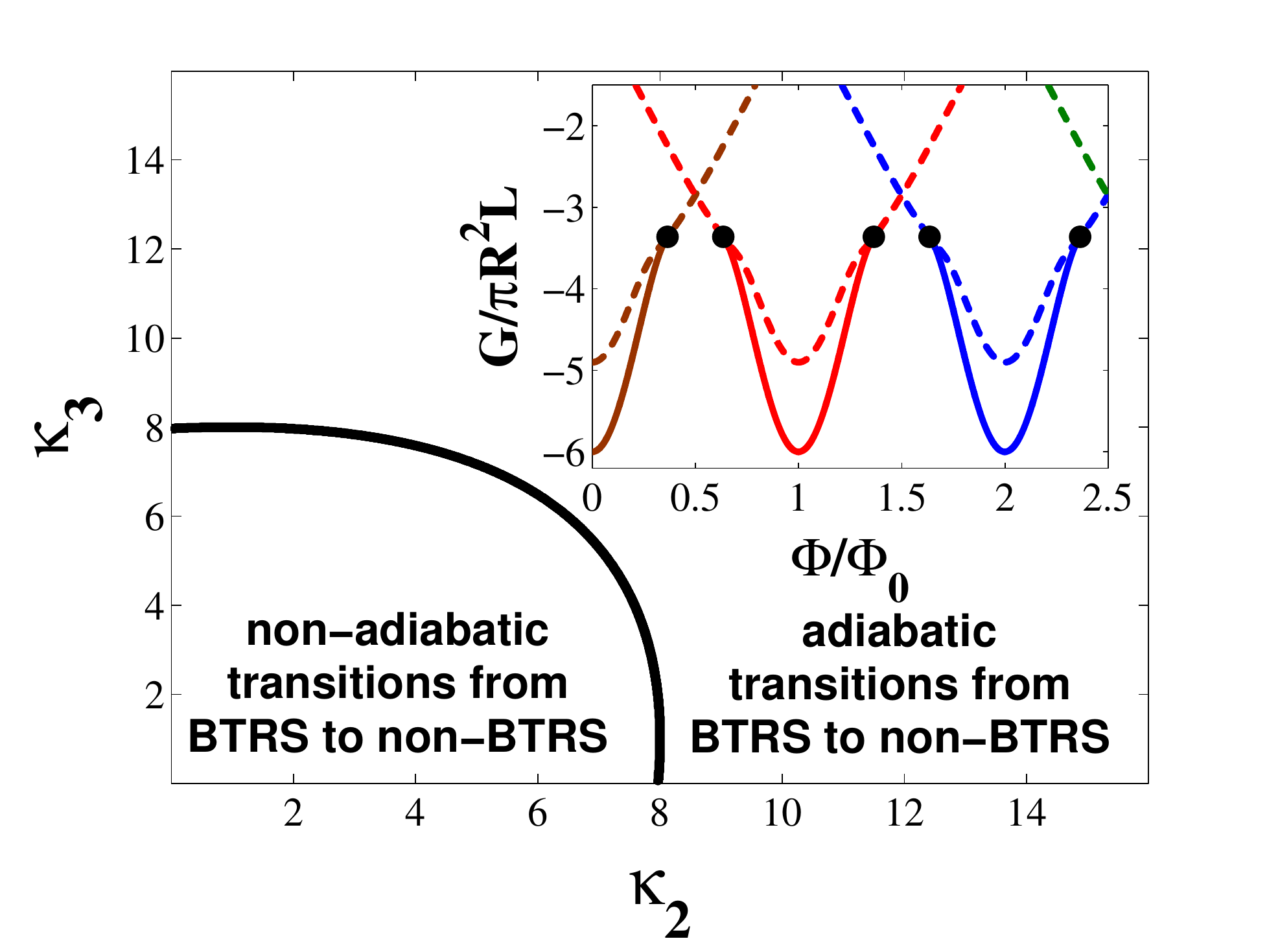}
\caption{Phase diagram of a three-band SC, where transitions
from BTRS to TRS can
occur due to an
excitation by an external
magnetic flux in the experimental setup shown in Fig. 2]
  and without any
excitation along an adiabatic path of changing  the GS (pink region).
The inset shows the evolution
of the GL free energy  in dependence on
the applied magnetic flux for a three-band
superconductor with   $\tilde{\gamma }=1$  for  attractive interband
interactions between the first and the second bands, the first and the third bands
and a repulsive one between the second and the third bands,  and for $\kappa_2 =15$  and $\kappa_3 =1.5$.
Black circles: critical (final) points for BTRS states (solid lines). }
\label{fig.4}
\end{figure}

To demonstrate the induced transition from a BTRS to a TRS state by an external  magnetic field we consider a particular case of equal $a_i =a$, $b_i =b$,  but keeping the $\kappa_i$
different and $\gamma_{12}=\gamma_{23}= -\gamma_{13} =\gamma$. At zero $q$
a doubly degenerate BTRS state with $\phi=2\pi-\theta = 5\pi/3$ and
$\phi=2\pi-\theta = \pi/3$ is realized. With increase $q$ first at $q_c$,
which is the solution of the equation $\cos\phi_c,\theta_c =1$, we get
transitions to TRS states (see Supplement). In the TRS state
the minimization of the GL energy can not be done analytically.
In this
case a numerical procedure must be
applied.
The dependence of the GL
 free energy Eq.(\ref{eq.GLinMagneticField}) on the applied magnetic flux for different
 ratios of $\kappa_i$ is presented in
 Fig.~\ref{fig.4}.
 We
 track the evolution  with magnetic flux of
 one of the ground states, namely for
 $\phi =5\pi /3$, $\theta =\pi /3$.  The full procedure can be found in the Supplement.
We find that for a given value of $\tilde{\gamma }$
the ground state of a three-band superconductor
under consideration exhibits always a
BTRS.
This means that despite the value of the trapped flux,
 by increasing the flux we will move along the bottom part of the solid curves
 (Fig.\ref{fig.5}), following the ''route'' $l0-l1-l2-...$ But for an non-adiabatic,
 fast switched on  magnetic flux, the three-band superconducting system
 can be excited and can be flipped to a
metastable states with TRS. For instance, the previous ground state ''route'' $l0-l1-l2-l3-...$ can be replaced by the path $l0-l1-l1'-l2'-l2-l3-...$, where the dashed part  $l1'-l2'$ corresponds to the mentioned above metastable state with TRS of the three-band superconducting tube, or to a more complicated ``route'', which will involve more excited states with TRS. Also we found that if $q_{c} <1/2$
then the transitions between BTRS and TRS states can occur without any excitation by an external magnetic
flux.  This means that solid (BTRS) and dashed (TRS) lines cross before $\Phi/\Phi_0=1/2$ (see inset in Fig. 5). The phase
diagram for a three-band superconductor, which determines the intervals of the
parameters $\kappa_{i}$  for the transitions from a  BTRS to a TRS state with an
excitation and without one is given in Fig. \ref{fig.4}.

Further numerical examination
 give that, if $\kappa_{i} >1$ ($i\ge 2$) , a non-adiabatic switching on
 the increase of the magnetic flux can lead to
 a transformation of a three-band SC with a BTRS GS
 into an excited state with  TRS
 and an $s_{+++}$ order parameter (see Fig. 6a and 6c)
 and then it relaxes again to a BTRS GS.  If one or
 both $\kappa_{i} <1$ ($i\ge 2$), then the increasing magnetic flux can transform
 a three-band SC with BTRS into
 an $s_{+\pm}$ three-band SC and finally again to a BTRS state (see Fig. 6b).
\begin{figure}
\includegraphics[width=0.49\columnwidth]{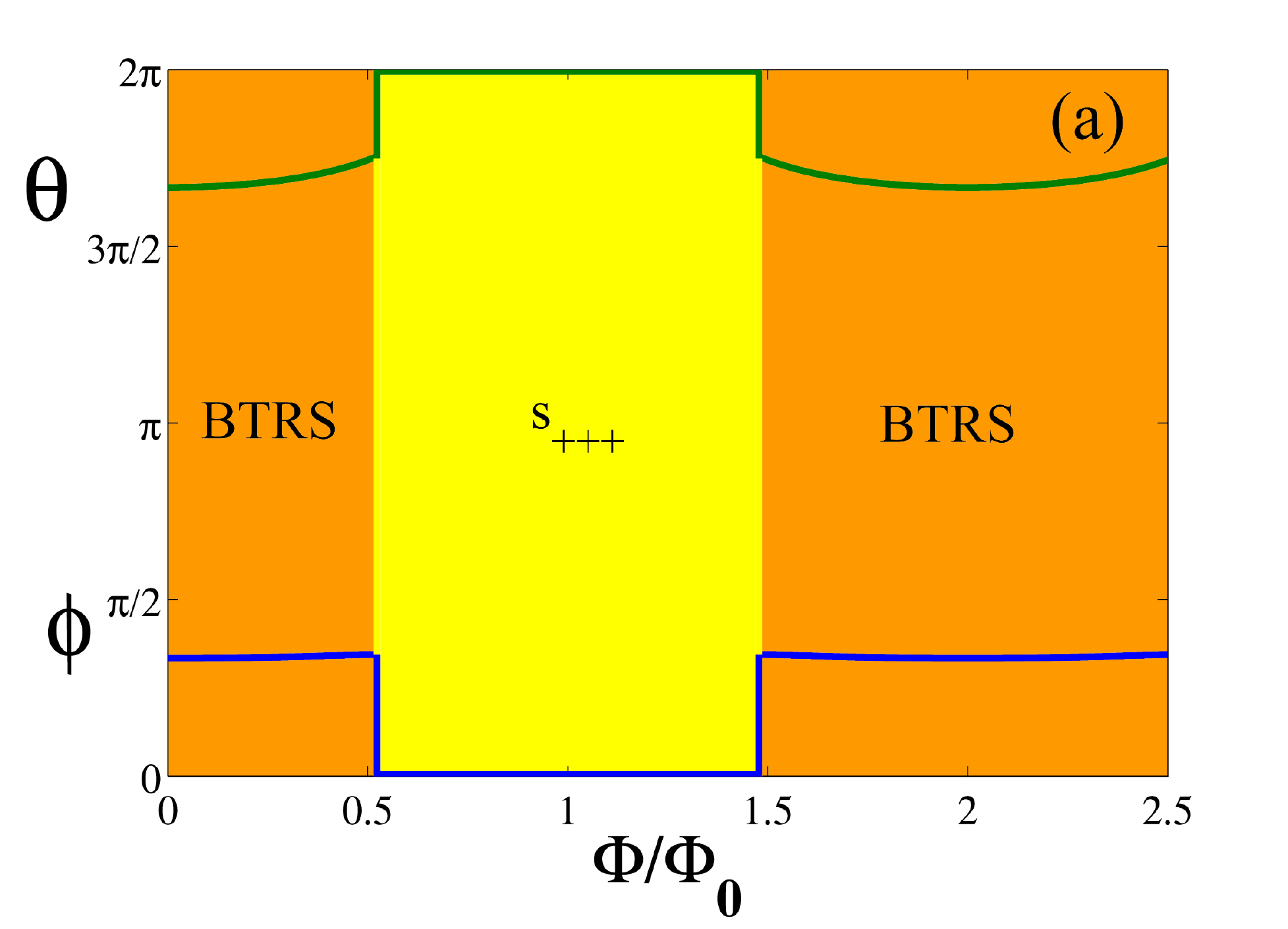}
\includegraphics[width=0.49\columnwidth]{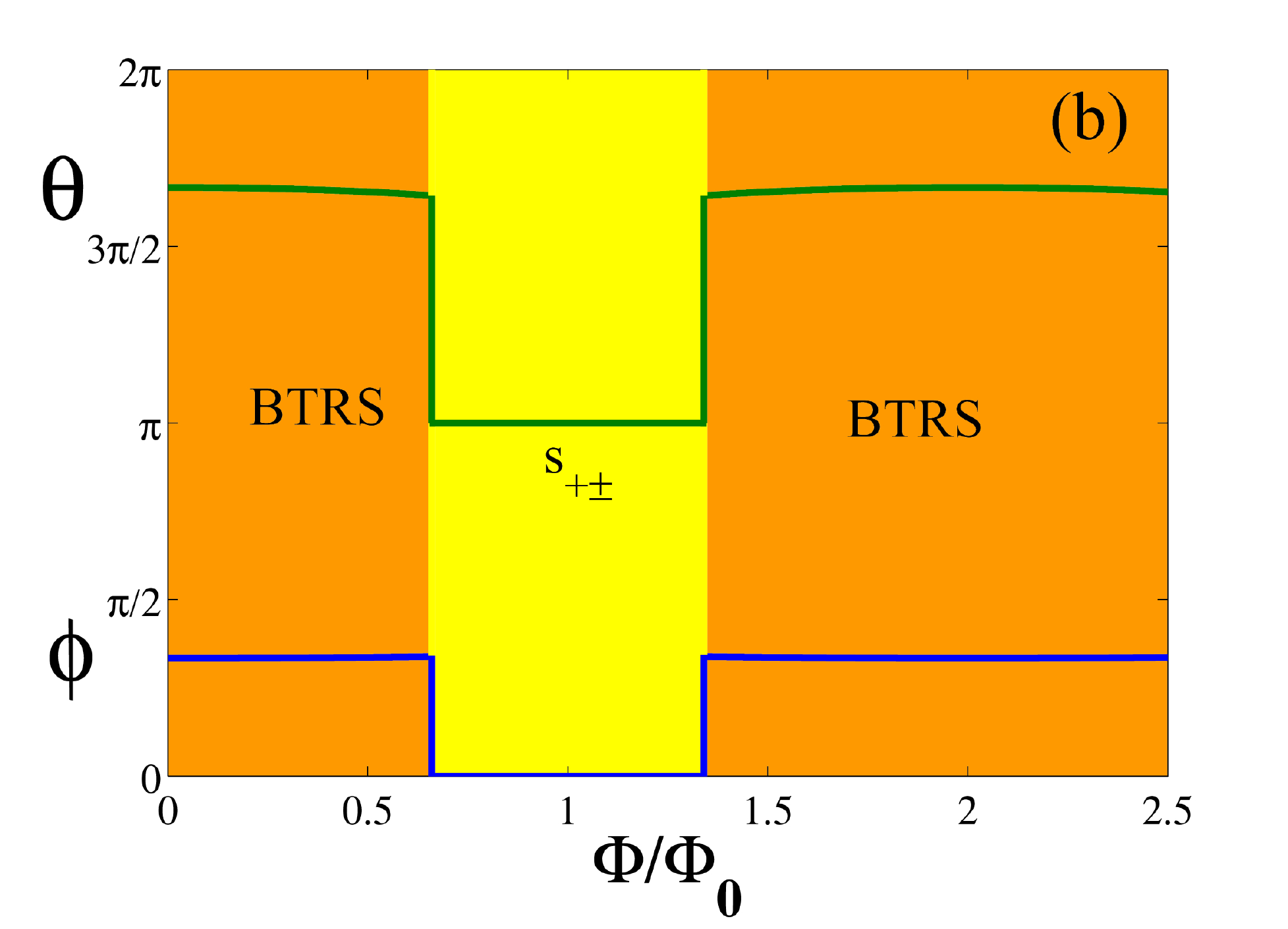}
\includegraphics[width=0.49\columnwidth]{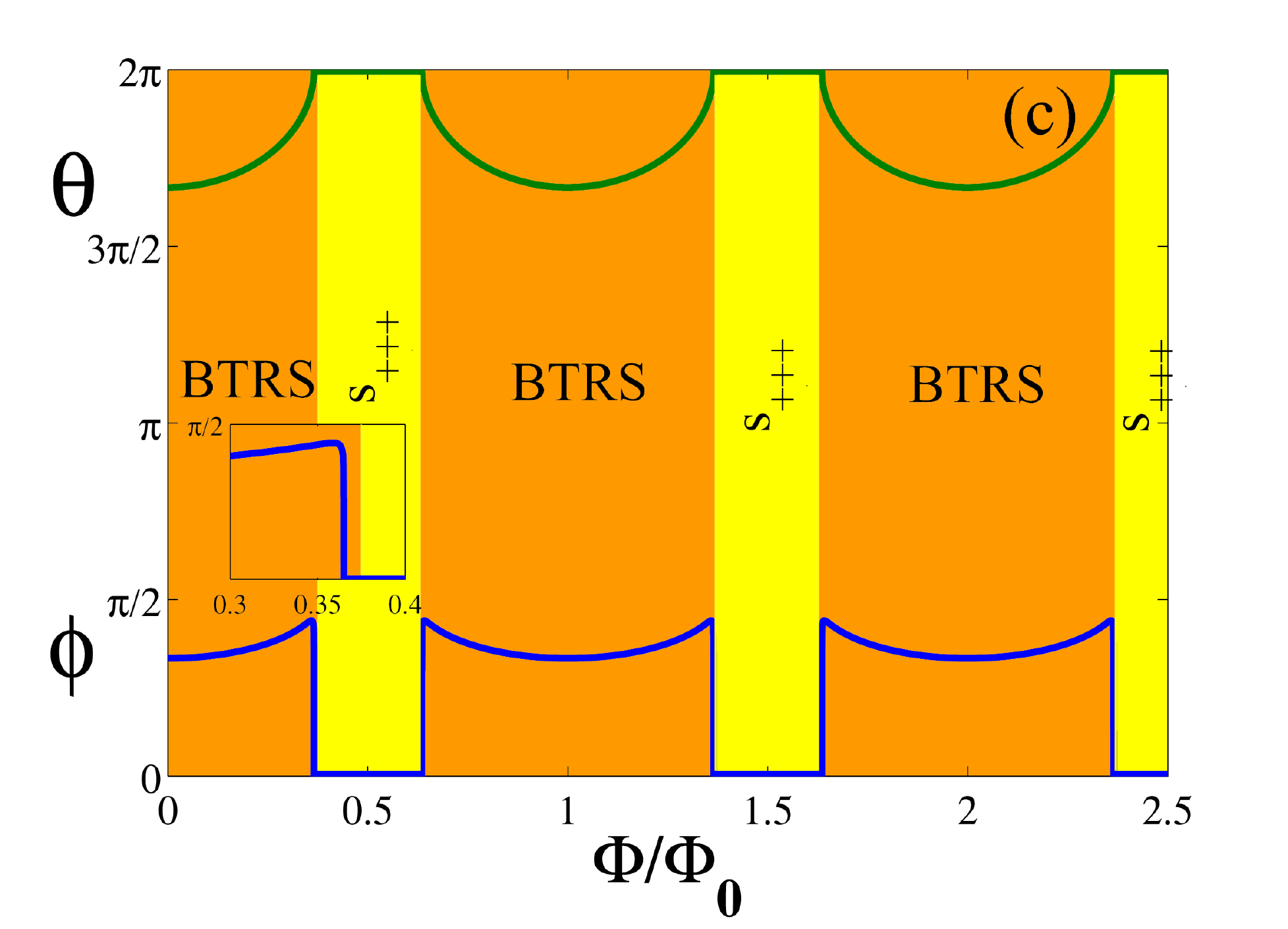}
\caption{(Color online) Possible evolution of the phase differences $\phi $
(blue) and $\theta $  (green) for a three-band superconductor with
non-adiabatic (a, b) and adiabatic (c) transitions between
BTRS and TRS states. The parameters are $\kappa_2 =4$  and $\kappa_3 =2$  (a),
$\kappa_2 =0.25$  and  $\kappa_3 =0.5$  (b) and $\kappa_2 =15$  and $\kappa_3 =1.5$ (c).
}
\label{fig.6}
\end{figure}

\begin{figure}
\includegraphics[width=0.49\columnwidth]{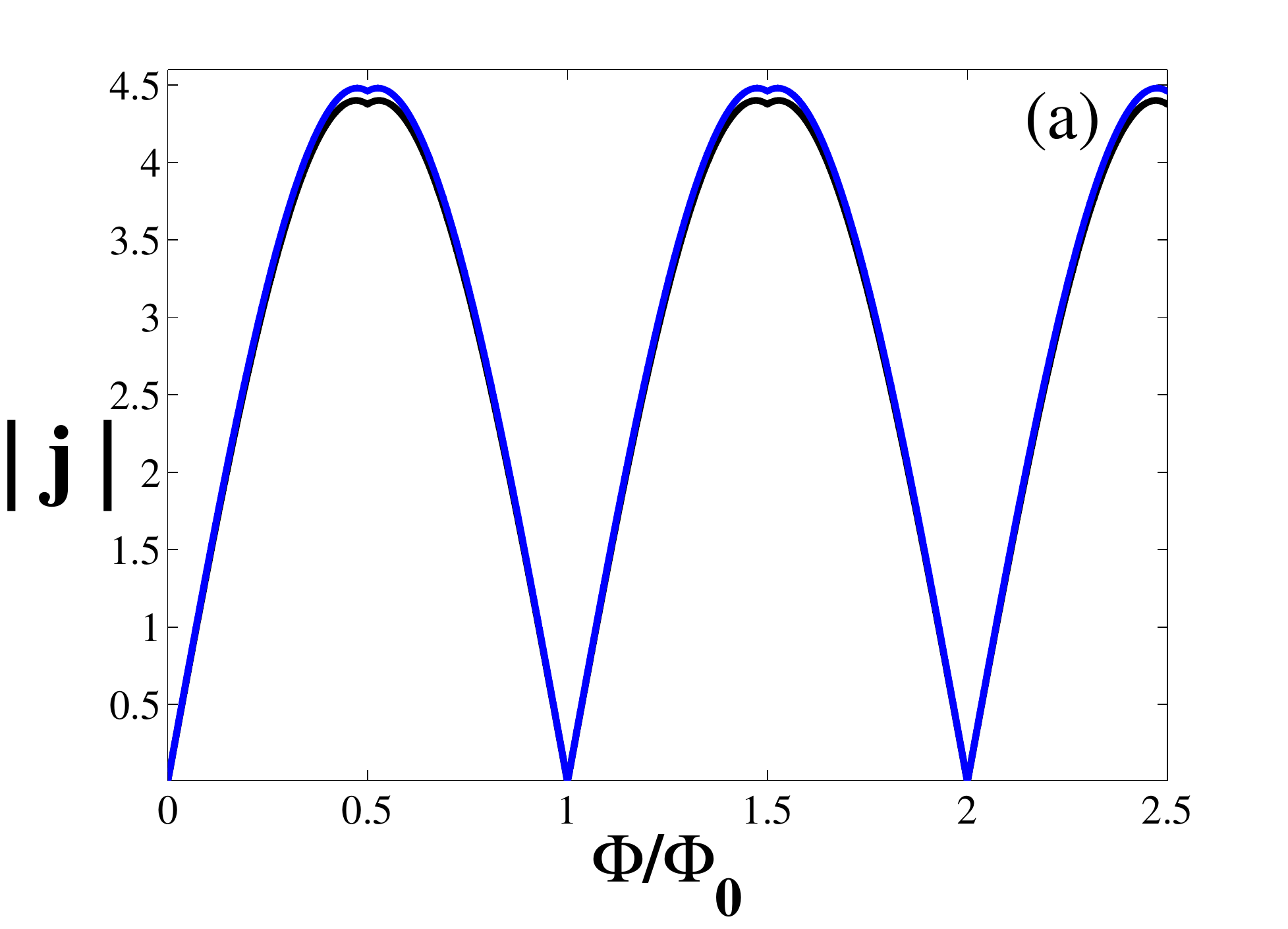}
\includegraphics[width=0.49\columnwidth]{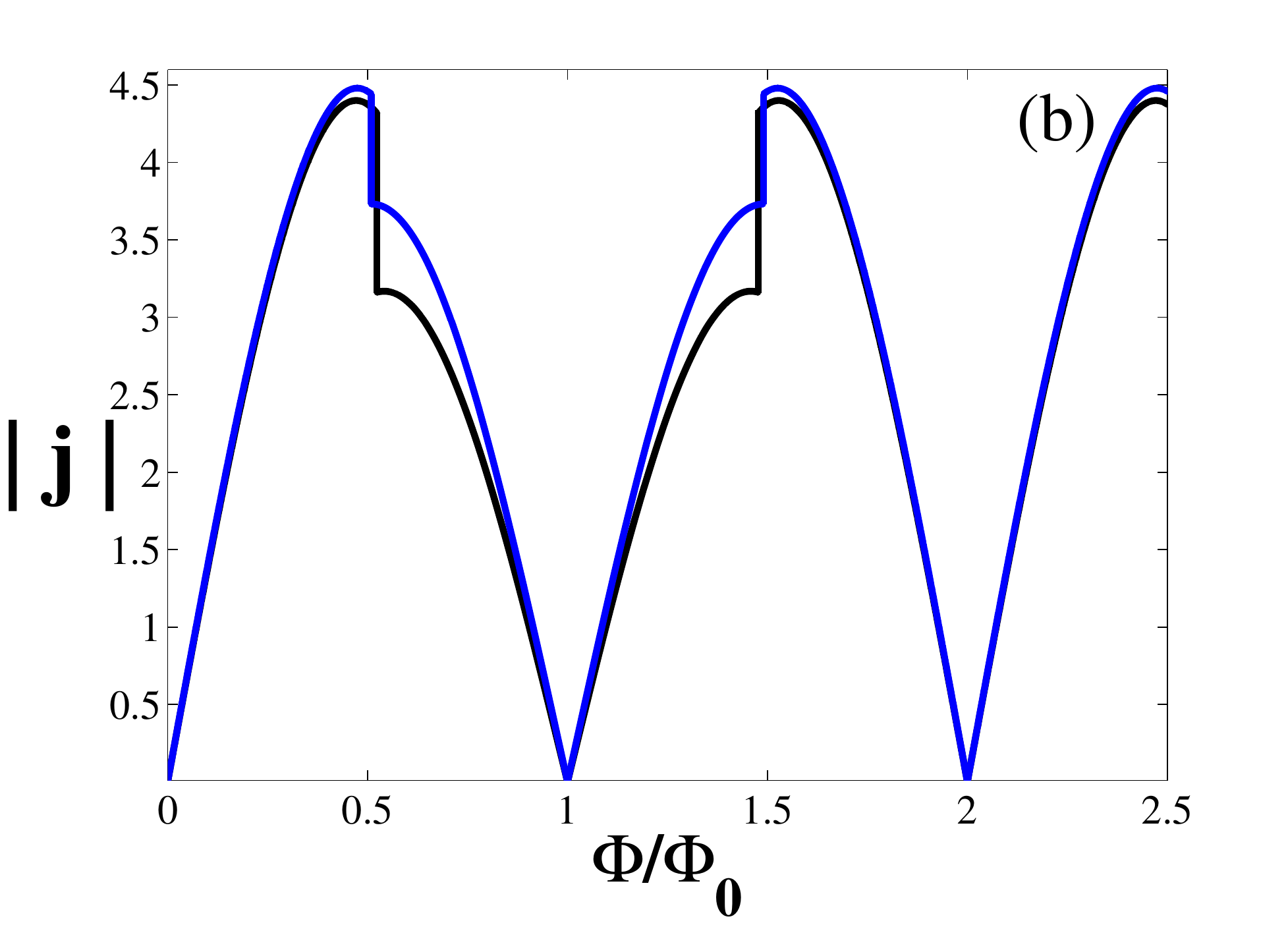}
\includegraphics[width=0.49\columnwidth]{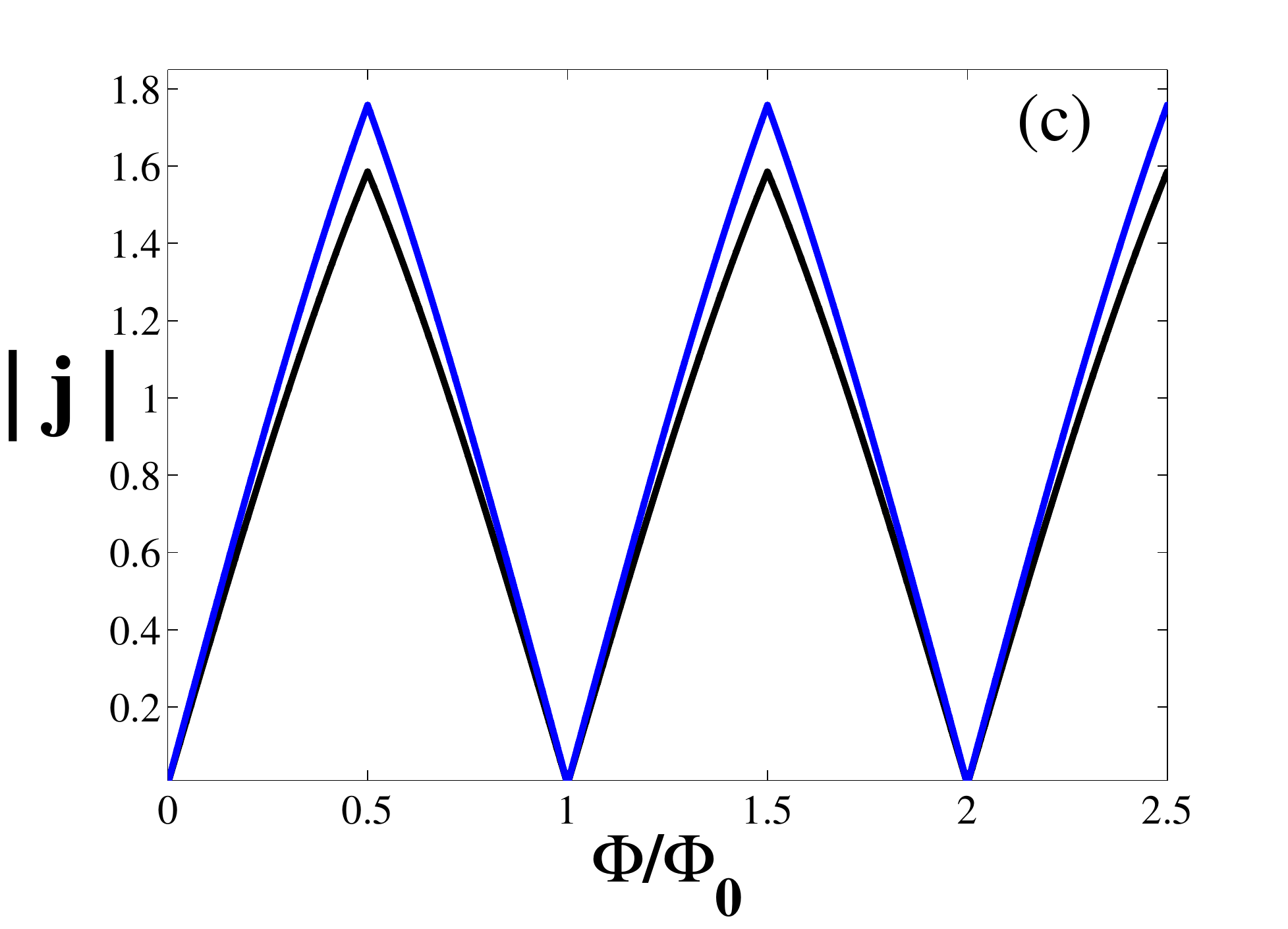}
\includegraphics[width=0.49\columnwidth]{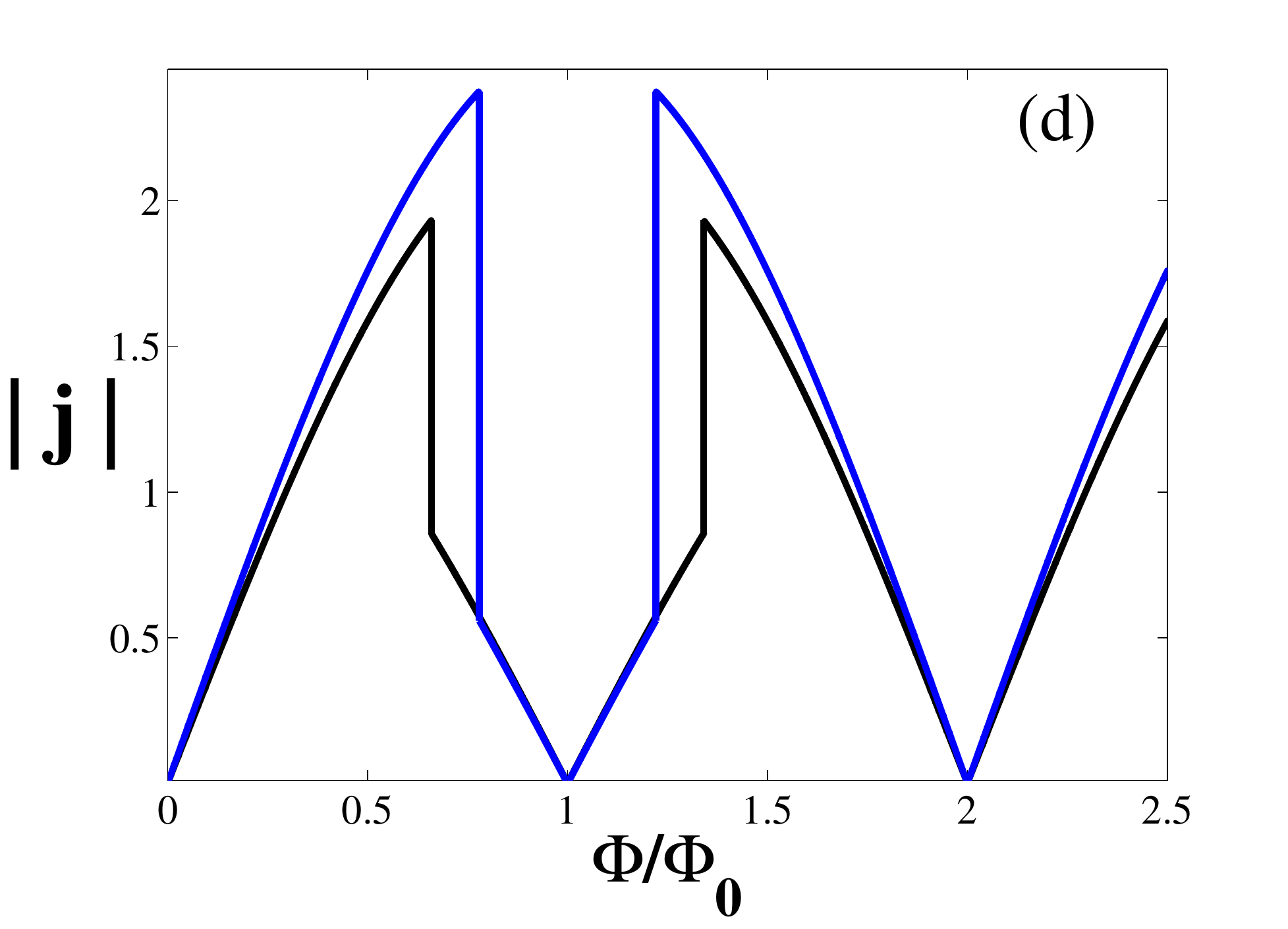}
\includegraphics[width=0.49\columnwidth]{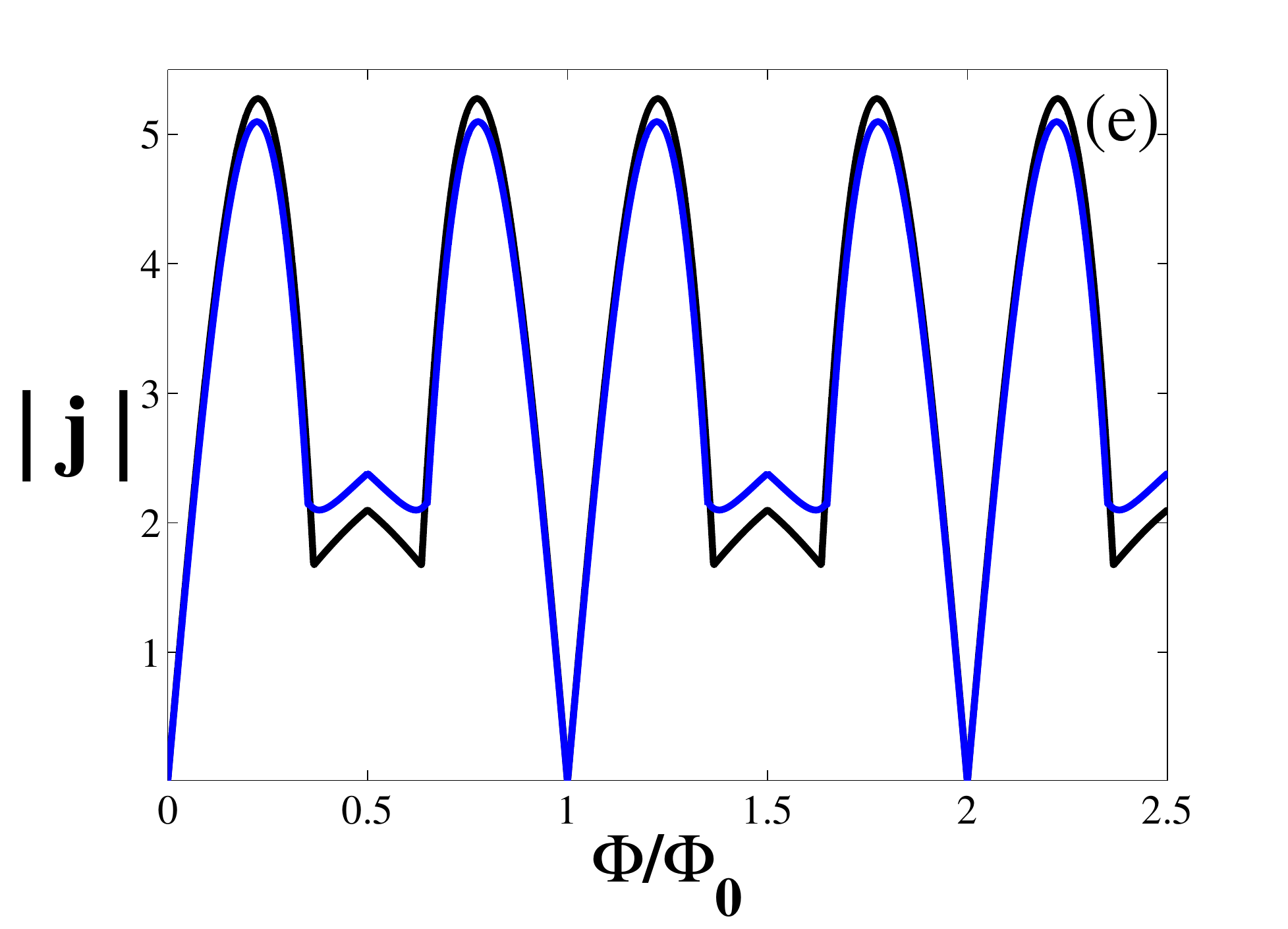}
\caption{(Color online) Current densities vs.\ the applied magnetic
flux in the setup shown in Fig.\ \ref{fig.tube}
made employing a three-band SC
with $\gamma_{12}=1$, $\gamma_{23}=-1$, $\gamma_{13}=1$ (black line) and with
$\gamma_{12}=1.1$, $\gamma_{23}=-1$, $\gamma_{13}=1.2$ (blue line) and
for $\kappa_2 =4$  and
 $\kappa_3 =2$  (a, b) $\kappa_2 =0.25$  and  $\kappa_3 =0.5$  (c, d) and
 $\kappa_2 =15$  and  $\kappa_3 =1.5$ (e).
 The curves (a) and (c) correspond to a three-band SC without
 transitions between BTRS and TRS states. The plots
 (b), (d) and (e) are for a three-band SC with transitions between
 BTRS and TRS states with and without excitation by an external
 magnetic field, respectively.}
 \label{fig.7}
\end{figure}

From the experimental point of view transitions from  BTRS state to
TRS ones and vice versa can be detected following the response of
the current
density on an applied magnetic flux (see Fig.\ 7). We revealed appropriate jumps on
the $j\left(\Phi /\Phi _{0} \right)$ dependencies (see Figs. 7b, 7d and 7e)
induced by these transitions.
Since in the present geometry changing of the magnetic field is equivalent to changing
temperature, we expect also special features for the specific heat
related to these BTRS to TRS transition. Hence accompanying thermodynamic measurements
might provide further support for identification of the BTRS states.
\section{Discussions and Conclusions}
Based on these analytical and numerical calculations
it is natural to suggest
that such a
behavior remains on a qualitative level the same also for other possible sets of interband interaction coefficients which admit the existence of frustrated states
in the equilibrium state (see supplemental materials). \cite{remark}
In other words jumps in the current dependencies in the magnetic flux driven regime can be expected for any three-band superconductor with a primordial (before switching on a magnetic field) BTRS state. Moreover with some restrictions it is reasonable to expect the same behavior
also for other BTRS multi-band superconductors, whose electronic structure
and physical properties are described by more than three order parameters
and where frustrated states are global ground states. Restrictions of the
application of such method are connected with the special case of  multi-band
superconductors with even number of bands and all equal repulsive interband
interactions, where BTRS and TRS states have the same energy. \cite{Tanaka2011,Yanisagawa2013,Weston2013}
We believe that the presence of such jumps can be considered as an
experimental proof for the detection of BTRS and frustration in unconventional
three- and multi-band superconductors.
It is important to note that the detection method proposed here compares favorably with
surface-sensitive techniques (interference or proximity based contacts)
for the detection of
properties related to the symmetry
of the order parameter, because
it probes the entire volume of the superconductor under examination.
Based on our results we propose to detect the presence of frustration and
BTRS in experiments with mesoscopic thin rings or tubes made from unconventional
three-band superconductors, by
measuring a generic current response on the
applied magnetic flux.

It should be noted that currently the exact location of the soliton
states on the energetic scale of a three-band superconductor is not
known. 
 Knowledge of all possible topological defects and their energies
  in case of three- and other multi-band superconductors is very important
  for the detection of the BTRS phenomenon in order to distinguish the jumps,
  connected with the presence of BTRS to TRS transitions and from
  the transitions from a  BTRS ground state to excited soliton states.
If the energy of phase-inhomogeneous solutions is higher than the  BTRS
and TRS states then during the excitation one can in principle
observe additional jumps on the current-magnetic flux dependencies due
to relaxation processes from higher energetic levels (soliton states)
to the ground state via metastable TRS states.
Another situation is realized for solitons, whose energy is within the
interval between BTRS and TRS states. In this case during the
excitation process a three-band superconductor can be promoted to a
TRS state as an intermediate state and then relax to the
ground state via other intermediate states of  solitonic nature. So also in this case additional jumps will also appear on the
experimental  dependencies.
The last possibility can occur if the BTRS state is not a globally stable
state and the ground state of a three-band superconductor already contains 
solitons. The realization of such a scenario was predicted recently for
a three-band superconductor \cite{Lin2012}
based on non-rigorous stability considerations
of phase kinks for an infinitely extended
superconducting system.
To the best of our knowledge a study of topological
defects in three-band superconductors for a  doubly-connected finite
superconducting system as considered here is still lacking. And it is not clear whether such solitons in a
restricted geometry can also occur as globally stable phenomena. We will study
this interesting but complex problem in more detail in the future.
Also the cases of imperfect three-band superconductors with impurities as well as inhomogeneous states due to the presence of solitonic nonlinear excitations mentioned above requires a special analysis outside of the scope of the present paper and be left for future study.

%
%

\section{Acknowledgment}
We acknowledge M.\ Kiselev, O.\ Dolgov,
M.\ Zhitomirsky, N.M.\ Plakida J.\ Schmalian, and A.\ Chubukov for helpful and critical discussions. Y.Y.\ thanks the ITF at the IFW-Dresden for hospitality and financial support where large parts of the present work has been performed. D.E.\ and S.-L.D.\ acknowledge the VW-foundation for partial support within the trilateral grant "{\it Synthesis, theoretical examination and experimental investigation of emergent iron-based superconductors}". This work was supported by the DFG-RSF grant and SFB1143. Y.Y. is grateful for support by the Russian Scientific Foundation Grant No 15-12-10020.

\bibliographystyle{apsrev}

\bibliography{CurrentsInSc}

\begin{thebibliography}{48}
\expandafter\ifx\csname natexlab\endcsname\relax\def\natexlab#1{#1}\fi
\expandafter\ifx\csname bibnamefont\endcsname\relax
  \def\bibnamefont#1{#1}\fi
\expandafter\ifx\csname bibfnamefont\endcsname\relax
  \def\bibfnamefont#1{#1}\fi
\expandafter\ifx\csname citenamefont\endcsname\relax
  \def\citenamefont#1{#1}\fi
\expandafter\ifx\csname url\endcsname\relax
  \def\url#1{\texttt{#1}}\fi
\expandafter\ifx\csname urlprefix\endcsname\relax\def\urlprefix{URL }\fi
\providecommand{\bibinfo}[2]{#2}
\providecommand{\eprint}[2][]{\url{#2}}

\bibitem[{\citenamefont{Marciani et~al.}(2013)\citenamefont{Marciani,
  Fanfarillo, Castellani, and Benfatto}}]{Marciani2013}
\bibinfo{author}{\bibfnamefont{M.}~\bibnamefont{Marciani}},
  \bibinfo{author}{\bibfnamefont{L.}~\bibnamefont{Fanfarillo}},
  \bibinfo{author}{\bibfnamefont{C.}~\bibnamefont{Castellani}},
  \bibnamefont{and} \bibinfo{author}{\bibfnamefont{L.}~\bibnamefont{Benfatto}},
  \bibinfo{journal}{Phys.\ Rev.\ B} \textbf{\bibinfo{volume}{88}},
  \bibinfo{pages}{214508} (\bibinfo{year}{2013}).

\bibitem[{\citenamefont{Carlstrom et~al.}(2011)\citenamefont{Carlstrom, Garaud,
  and Babaev}}]{Carlstrom2011}
\bibinfo{author}{\bibfnamefont{J.}~\bibnamefont{Carlstrom}},
  \bibinfo{author}{\bibfnamefont{J.}~\bibnamefont{Garaud}}, \bibnamefont{and}
  \bibinfo{author}{\bibfnamefont{E.}~\bibnamefont{Babaev}},
  \bibinfo{journal}{Phys.\ Rev.\ Lett.} \textbf{\bibinfo{volume}{107}},
  \bibinfo{pages}{197001} (\bibinfo{year}{2011}).

\bibitem[{\citenamefont{Bojesen et~al.}(2013)\citenamefont{Bojesen, Babaev, and
  Sudb\o{}}}]{Bojesen2013}
\bibinfo{author}{\bibfnamefont{T.}~\bibnamefont{Bojesen}},
  \bibinfo{author}{\bibfnamefont{E.}~\bibnamefont{Babaev}}, \bibnamefont{and}
  \bibinfo{author}{\bibfnamefont{A.}~\bibnamefont{Sudb\o{}}},
  \bibinfo{journal}{Phys. Rev. B} \textbf{\bibinfo{volume}{88}},
  \bibinfo{pages}{220511} (\bibinfo{year}{2013}).

\bibitem[{\citenamefont{Lin and Hu}(2012{\natexlab{a}})}]{Shi-Zheng2012}
\bibinfo{author}{\bibfnamefont{S.}~\bibnamefont{Lin}} \bibnamefont{and}
  \bibinfo{author}{\bibfnamefont{X.}~\bibnamefont{Hu}}, \bibinfo{journal}{N.\
  J.\ of Phys.} \textbf{\bibinfo{volume}{14}}, \bibinfo{pages}{063021}
  (\bibinfo{year}{2012}{\natexlab{a}}).

\bibitem[{\citenamefont{Yanagisawa et~al.}(2012)\citenamefont{Yanagisawa,
  Tanaka, Hase, and Yamaji}}]{Yanagisawa2012}
\bibinfo{author}{\bibfnamefont{T.}~\bibnamefont{Yanagisawa}},
  \bibinfo{author}{\bibfnamefont{Y.}~\bibnamefont{Tanaka}},
  \bibinfo{author}{\bibfnamefont{I.}~\bibnamefont{Hase}}, \bibnamefont{and}
  \bibinfo{author}{\bibfnamefont{K.}~\bibnamefont{Yamaji}},
  \bibinfo{journal}{J.\ Phys.\ Soc.\ Jpn.} \textbf{\bibinfo{volume}{81}},
  \bibinfo{pages}{024712} (\bibinfo{year}{2012}).

\bibitem[{\citenamefont{Lin}(2014)}]{Lin2014}
\bibinfo{author}{\bibfnamefont{S.-Z.} \bibnamefont{Lin}},
  \bibinfo{journal}{Journal of Physics: Condensed Matter}
  \textbf{\bibinfo{volume}{26}}, \bibinfo{pages}{493202}
  (\bibinfo{year}{2014}),
  \urlprefix\url{http://stacks.iop.org/0953-8984/26/i=49/a=493202}.

\bibitem[{\citenamefont{Tanaka}(2015)}]{Tanaka2015}
\bibinfo{author}{\bibfnamefont{Y.}~\bibnamefont{Tanaka}},
  \bibinfo{journal}{Superconductor Science and Technology}
  \textbf{\bibinfo{volume}{28}}, \bibinfo{pages}{034002}
  (\bibinfo{year}{2015}),
  \urlprefix\url{http://stacks.iop.org/0953-2048/28/i=3/a=034002}.

\bibitem[{\citenamefont{Lin et~al.}(2016)\citenamefont{Lin, Maiti, and
  Chubukov}}]{Chubukov2016}
\bibinfo{author}{\bibfnamefont{S.-Z.} \bibnamefont{Lin}},
  \bibinfo{author}{\bibfnamefont{S.}~\bibnamefont{Maiti}}, \bibnamefont{and}
  \bibinfo{author}{\bibfnamefont{A.}~\bibnamefont{Chubukov}},
  \bibinfo{journal}{Phys. Rev. B} \textbf{\bibinfo{volume}{94}},
  \bibinfo{pages}{064519} (\bibinfo{year}{2016}),
  \urlprefix\url{http://link.aps.org/doi/10.1103/PhysRevB.94.064519}.

\bibitem[{\citenamefont{Huang and Hu}(2016)}]{Huang2016}
\bibinfo{author}{\bibfnamefont{Z.}~\bibnamefont{Huang}} \bibnamefont{and}
  \bibinfo{author}{\bibfnamefont{X.}~\bibnamefont{Hu}},
  \bibinfo{journal}{Journal of Superconductivity and Novel Magnetism}
  \textbf{\bibinfo{volume}{29}}, \bibinfo{pages}{597} (\bibinfo{year}{2016}),
  ISSN \bibinfo{issn}{1557-1947},
  \urlprefix\url{http://dx.doi.org/10.1007/s10948-015-3309-x}.

\bibitem[{\citenamefont{Garaud et~al.}(2016)\citenamefont{Garaud, Silaev, and
  Babaev}}]{Garaud2016}
\bibinfo{author}{\bibfnamefont{J.}~\bibnamefont{Garaud}},
  \bibinfo{author}{\bibfnamefont{M.}~\bibnamefont{Silaev}}, \bibnamefont{and}
  \bibinfo{author}{\bibfnamefont{E.}~\bibnamefont{Babaev}},
  \bibinfo{journal}{Phys. Rev. Lett.} \textbf{\bibinfo{volume}{116}},
  \bibinfo{pages}{097002} (\bibinfo{year}{2016}),
  \urlprefix\url{http://link.aps.org/doi/10.1103/PhysRevLett.116.097002}.

\bibitem[{\citenamefont{Koyama}(2016)}]{Koyama2016}
\bibinfo{author}{\bibfnamefont{T.}~\bibnamefont{Koyama}},
  \bibinfo{journal}{Journal of the Physical Society of Japan}
  \textbf{\bibinfo{volume}{85}}, \bibinfo{pages}{064715}
  (\bibinfo{year}{2016}), \eprint{http://dx.doi.org/10.7566/JPSJ.85.064715},
  \urlprefix\url{http://dx.doi.org/10.7566/JPSJ.85.064715}.

\bibitem[{\citenamefont{Stanev}(2015)}]{Stanev2015}
\bibinfo{author}{\bibfnamefont{V.}~\bibnamefont{Stanev}},
  \bibinfo{journal}{Superconductor Science and Technology}
  \textbf{\bibinfo{volume}{28}}, \bibinfo{pages}{014006}
  (\bibinfo{year}{2015}),
  \urlprefix\url{http://stacks.iop.org/0953-2048/28/i=1/a=014006}.

\bibitem[{\citenamefont{Mackenzie and Maeno}(2003)}]{Mackenzie2003}
\bibinfo{author}{\bibfnamefont{A.~P.} \bibnamefont{Mackenzie}}
  \bibnamefont{and} \bibinfo{author}{\bibfnamefont{Y.}~\bibnamefont{Maeno}},
  \bibinfo{journal}{Rev.\ Mod. Phys.} \textbf{\bibinfo{volume}{75}},
  \bibinfo{pages}{657} (\bibinfo{year}{2003}).

\bibitem[{\citenamefont{Luke et~al.}(1998)\citenamefont{Luke, Fudamoto, Kojima,
  Larkin, Merrin, Nachumi, Uemura, Maeno, Mao, Mori et~al.}}]{luketime1998}
\bibinfo{author}{\bibfnamefont{G.}~\bibnamefont{Luke}},
  \bibinfo{author}{\bibfnamefont{Y.}~\bibnamefont{Fudamoto}},
  \bibinfo{author}{\bibfnamefont{K.}~\bibnamefont{Kojima}},
  \bibinfo{author}{\bibfnamefont{M.}~\bibnamefont{Larkin}},
  \bibinfo{author}{\bibfnamefont{J.}~\bibnamefont{Merrin}},
  \bibinfo{author}{\bibfnamefont{B.}~\bibnamefont{Nachumi}},
  \bibinfo{author}{\bibfnamefont{Y.}~\bibnamefont{Uemura}},
  \bibinfo{author}{\bibfnamefont{Y.}~\bibnamefont{Maeno}},
  \bibinfo{author}{\bibfnamefont{Z.}~\bibnamefont{Mao}},
  \bibinfo{author}{\bibfnamefont{Y.}~\bibnamefont{Mori}}, \bibnamefont{et~al.},
  \bibinfo{journal}{Nature} \textbf{\bibinfo{volume}{394}},
  \bibinfo{pages}{558} (\bibinfo{year}{1998}).

\bibitem[{\citenamefont{Heffner et~al.}(1990)\citenamefont{Heffner, Smith,
  Willis, Birrer, Baines, Gygax, Hitti, Lippelt, Ott, Schenck
  et~al.}}]{Heffneret}
\bibinfo{author}{\bibfnamefont{R.~H.} \bibnamefont{Heffner}},
  \bibinfo{author}{\bibfnamefont{J.~L.} \bibnamefont{Smith}},
  \bibinfo{author}{\bibfnamefont{J.~O.} \bibnamefont{Willis}},
  \bibinfo{author}{\bibfnamefont{P.}~\bibnamefont{Birrer}},
  \bibinfo{author}{\bibfnamefont{C.}~\bibnamefont{Baines}},
  \bibinfo{author}{\bibfnamefont{F.~N.} \bibnamefont{Gygax}},
  \bibinfo{author}{\bibfnamefont{B.}~\bibnamefont{Hitti}},
  \bibinfo{author}{\bibfnamefont{E.}~\bibnamefont{Lippelt}},
  \bibinfo{author}{\bibfnamefont{H.~R.} \bibnamefont{Ott}},
  \bibinfo{author}{\bibfnamefont{A.}~\bibnamefont{Schenck}},
  \bibnamefont{et~al.}, \bibinfo{journal}{Phys. Rev. Lett.}
  \textbf{\bibinfo{volume}{65}}, \bibinfo{pages}{2816} (\bibinfo{year}{1990}),
  \urlprefix\url{http://link.aps.org/doi/10.1103/PhysRevLett.65.2816}.

\bibitem[{\citenamefont{Luke et~al.}(1993)\citenamefont{Luke, Keren, Le, Wu,
  Uemura, Bonn, Taillefer, and Garrett}}]{Lukeet}
\bibinfo{author}{\bibfnamefont{G.~M.} \bibnamefont{Luke}},
  \bibinfo{author}{\bibfnamefont{A.}~\bibnamefont{Keren}},
  \bibinfo{author}{\bibfnamefont{L.~P.} \bibnamefont{Le}},
  \bibinfo{author}{\bibfnamefont{W.~D.} \bibnamefont{Wu}},
  \bibinfo{author}{\bibfnamefont{Y.~J.} \bibnamefont{Uemura}},
  \bibinfo{author}{\bibfnamefont{D.~A.} \bibnamefont{Bonn}},
  \bibinfo{author}{\bibfnamefont{L.}~\bibnamefont{Taillefer}},
  \bibnamefont{and} \bibinfo{author}{\bibfnamefont{J.~D.}
  \bibnamefont{Garrett}}, \bibinfo{journal}{Phys. Rev. Lett.}
  \textbf{\bibinfo{volume}{71}}, \bibinfo{pages}{1466} (\bibinfo{year}{1993}),
  \urlprefix\url{http://link.aps.org/doi/10.1103/PhysRevLett.71.1466}.

\bibitem[{\citenamefont{Biswas et~al.}(2013)\citenamefont{Biswas, Luetkens,
  Neupert, St\"urzer, Baines, Pascua, Schnyder, Fischer, Goryo, Lees
  et~al.}}]{Biswas}
\bibinfo{author}{\bibfnamefont{P.~K.} \bibnamefont{Biswas}},
  \bibinfo{author}{\bibfnamefont{H.}~\bibnamefont{Luetkens}},
  \bibinfo{author}{\bibfnamefont{T.}~\bibnamefont{Neupert}},
  \bibinfo{author}{\bibfnamefont{T.}~\bibnamefont{St\"urzer}},
  \bibinfo{author}{\bibfnamefont{C.}~\bibnamefont{Baines}},
  \bibinfo{author}{\bibfnamefont{G.}~\bibnamefont{Pascua}},
  \bibinfo{author}{\bibfnamefont{A.~P.} \bibnamefont{Schnyder}},
  \bibinfo{author}{\bibfnamefont{M.~H.} \bibnamefont{Fischer}},
  \bibinfo{author}{\bibfnamefont{J.}~\bibnamefont{Goryo}},
  \bibinfo{author}{\bibfnamefont{M.~R.} \bibnamefont{Lees}},
  \bibnamefont{et~al.}, \bibinfo{journal}{Phys. Rev. B}
  \textbf{\bibinfo{volume}{87}}, \bibinfo{pages}{180503}
  (\bibinfo{year}{2013}),
  \urlprefix\url{http://link.aps.org/doi/10.1103/PhysRevB.87.180503}.

\bibitem[{\citenamefont{Laughlin}(1998)}]{Laughlin1998}
\bibinfo{author}{\bibfnamefont{R.~B.} \bibnamefont{Laughlin}},
  \bibinfo{journal}{Phys.\ Rev.\ Lett.} \textbf{\bibinfo{volume}{80}},
  \bibinfo{pages}{5188} (\bibinfo{year}{1998}).

\bibitem[{\citenamefont{Krishana et~al.}(1997)\citenamefont{Krishana, Ong, Li,
  Gu, and Koshizuka}}]{Krishana1997}
\bibinfo{author}{\bibfnamefont{K.}~\bibnamefont{Krishana}},
  \bibinfo{author}{\bibfnamefont{N.}~\bibnamefont{Ong}},
  \bibinfo{author}{\bibfnamefont{Q.}~\bibnamefont{Li}},
  \bibinfo{author}{\bibfnamefont{G.~D.} \bibnamefont{Gu}}, \bibnamefont{and}
  \bibinfo{author}{\bibfnamefont{N.}~\bibnamefont{Koshizuka}},
  \bibinfo{journal}{Science} \textbf{\bibinfo{volume}{277}},
  \bibinfo{pages}{83} (\bibinfo{year}{1997}), ISSN \bibinfo{issn}{0036-8075},
  \eprint{http://science.sciencemag.org/content/277/5322/83.full.pdf},
  \urlprefix\url{http://science.sciencemag.org/content/277/5322/83}.

\bibitem[{\citenamefont{Ganesh et~al.}(2014)\citenamefont{Ganesh, Baskaran,
  van~den Brink, and Efremov}}]{Efremov2014}
\bibinfo{author}{\bibfnamefont{R.}~\bibnamefont{Ganesh}},
  \bibinfo{author}{\bibfnamefont{G.}~\bibnamefont{Baskaran}},
  \bibinfo{author}{\bibfnamefont{J.}~\bibnamefont{van~den Brink}},
  \bibnamefont{and} \bibinfo{author}{\bibfnamefont{D.~V.}
  \bibnamefont{Efremov}}, \bibinfo{journal}{Phys. Rev. Lett.}
  \textbf{\bibinfo{volume}{113}}, \bibinfo{pages}{177001}
  (\bibinfo{year}{2014}),
  \urlprefix\url{http://link.aps.org/doi/10.1103/PhysRevLett.113.177001}.

\bibitem[{\citenamefont{Nandkishore et~al.}(2014)\citenamefont{Nandkishore,
  Thomale, and Chubukov}}]{Thomale2014}
\bibinfo{author}{\bibfnamefont{R.}~\bibnamefont{Nandkishore}},
  \bibinfo{author}{\bibfnamefont{R.}~\bibnamefont{Thomale}}, \bibnamefont{and}
  \bibinfo{author}{\bibfnamefont{A.}~\bibnamefont{Chubukov}},
  \bibinfo{journal}{Phys\. Rev.\ B} \textbf{\bibinfo{volume}{89}},
  \bibinfo{pages}{144501} (\bibinfo{year}{2014}).

\bibitem[{\citenamefont{Kiesel et~al.}(2013)\citenamefont{Kiesel, Platt, Hanke,
  and Thomale}}]{Thomale2013}
\bibinfo{author}{\bibfnamefont{M.}~\bibnamefont{Kiesel}},
  \bibinfo{author}{\bibfnamefont{C.}~\bibnamefont{Platt}},
  \bibinfo{author}{\bibfnamefont{W.}~\bibnamefont{Hanke}}, \bibnamefont{and}
  \bibinfo{author}{\bibfnamefont{R.}~\bibnamefont{Thomale}},
  \bibinfo{journal}{Phys.\ Rev.\ Lett.} \textbf{\bibinfo{volume}{111}},
  \bibinfo{pages}{097001} (\bibinfo{year}{2013}).

\bibitem[{\citenamefont{Nandkishore et~al.}(2012)\citenamefont{Nandkishore,
  Levitov, and Chubukov}}]{Nandkishore2012}
\bibinfo{author}{\bibfnamefont{R.}~\bibnamefont{Nandkishore}},
  \bibinfo{author}{\bibfnamefont{L.}~\bibnamefont{Levitov}}, \bibnamefont{and}
  \bibinfo{author}{\bibfnamefont{A.}~\bibnamefont{Chubukov}},
  \bibinfo{journal}{Nature Physics} \textbf{\bibinfo{volume}{8}},
  \bibinfo{pages}{158} (\bibinfo{year}{2012}).

\bibitem[{\citenamefont{Tafti et~al.}(2013)\citenamefont{Tafti, Juneau-Fecteau,
  Delage, Ren~de Cotret, Reid, Wang, Luo, Chen, Doiron-Leyraud, and
  Taillefer}}]{Tafti2013}
\bibinfo{author}{\bibfnamefont{F.}~\bibnamefont{Tafti}},
  \bibinfo{author}{\bibfnamefont{A.}~\bibnamefont{Juneau-Fecteau}},
  \bibinfo{author}{\bibfnamefont{M.}~\bibnamefont{Delage}},
  \bibinfo{author}{\bibfnamefont{S.}~\bibnamefont{Ren~de Cotret}},
  \bibinfo{author}{\bibfnamefont{J.-P.} \bibnamefont{Reid}},
  \bibinfo{author}{\bibfnamefont{A.}~\bibnamefont{Wang}},
  \bibinfo{author}{\bibfnamefont{X.-G.} \bibnamefont{Luo}},
  \bibinfo{author}{\bibfnamefont{X.}~\bibnamefont{Chen}},
  \bibinfo{author}{\bibfnamefont{N.}~\bibnamefont{Doiron-Leyraud}},
  \bibnamefont{and}
  \bibinfo{author}{\bibfnamefont{L.}~\bibnamefont{Taillefer}},
  \bibinfo{journal}{Nature Physics} \textbf{\bibinfo{volume}{9}},
  \bibinfo{pages}{349} (\bibinfo{year}{2013}).

\bibitem[{\citenamefont{Terashima et~al.}(2013)\citenamefont{Terashima, Kurita,
  Kimata, Tomita, Tsuchiya, Imai, Sato, Kihou, Lee, Kito
  et~al.}}]{Terashima2013}
\bibinfo{author}{\bibfnamefont{T.}~\bibnamefont{Terashima}},
  \bibinfo{author}{\bibfnamefont{N.}~\bibnamefont{Kurita}},
  \bibinfo{author}{\bibfnamefont{M.}~\bibnamefont{Kimata}},
  \bibinfo{author}{\bibfnamefont{M.}~\bibnamefont{Tomita}},
  \bibinfo{author}{\bibfnamefont{S.}~\bibnamefont{Tsuchiya}},
  \bibinfo{author}{\bibfnamefont{M.}~\bibnamefont{Imai}},
  \bibinfo{author}{\bibfnamefont{A.}~\bibnamefont{Sato}},
  \bibinfo{author}{\bibfnamefont{K.}~\bibnamefont{Kihou}},
  \bibinfo{author}{\bibfnamefont{C.-H.} \bibnamefont{Lee}},
  \bibinfo{author}{\bibfnamefont{H.}~\bibnamefont{Kito}}, \bibnamefont{et~al.},
  \bibinfo{journal}{Phys.\ Rev.\ B} \textbf{\bibinfo{volume}{87}},
  \bibinfo{pages}{224512} (\bibinfo{year}{2013}).

\bibitem[{\citenamefont{Abdel-Hafiez et~al.}(2013)\citenamefont{Abdel-Hafiez,
  Grinenko, Aswartham, Morozov, Roslova, Vakaliuk, Johnston, Efremov, van~den
  Brink, Rosner et~al.}}]{Abdel-Hafiez2013}
\bibinfo{author}{\bibfnamefont{M.}~\bibnamefont{Abdel-Hafiez}},
  \bibinfo{author}{\bibfnamefont{V.}~\bibnamefont{Grinenko}},
  \bibinfo{author}{\bibfnamefont{S.}~\bibnamefont{Aswartham}},
  \bibinfo{author}{\bibfnamefont{I.}~\bibnamefont{Morozov}},
  \bibinfo{author}{\bibfnamefont{M.}~\bibnamefont{Roslova}},
  \bibinfo{author}{\bibfnamefont{O.}~\bibnamefont{Vakaliuk}},
  \bibinfo{author}{\bibfnamefont{S.}~\bibnamefont{Johnston}},
  \bibinfo{author}{\bibfnamefont{D.}~\bibnamefont{Efremov}},
  \bibinfo{author}{\bibfnamefont{J.}~\bibnamefont{van~den Brink}},
  \bibinfo{author}{\bibfnamefont{H.}~\bibnamefont{Rosner}},
  \bibnamefont{et~al.}, \bibinfo{journal}{Phys.\ Rev.\ B}
  \textbf{\bibinfo{volume}{87}}, \bibinfo{pages}{180507(R)}
  (\bibinfo{year}{2013}).

\bibitem[{\citenamefont{Grinenko et~al.}(2014)\citenamefont{Grinenko, Efremov,
  Drechsler, Aswartham, D., M., Morozov, Nenkov, Wurmehl, Wolter
  et~al.}}]{Grinenko2014}
\bibinfo{author}{\bibfnamefont{V.}~\bibnamefont{Grinenko}},
  \bibinfo{author}{\bibfnamefont{D.}~\bibnamefont{Efremov}},
  \bibinfo{author}{\bibfnamefont{S.-L.} \bibnamefont{Drechsler}},
  \bibinfo{author}{\bibfnamefont{S.}~\bibnamefont{Aswartham}},
  \bibinfo{author}{\bibfnamefont{G.}~\bibnamefont{D.}},
  \bibinfo{author}{\bibfnamefont{R.}~\bibnamefont{M.}},
  \bibinfo{author}{\bibfnamefont{I.}~\bibnamefont{Morozov}},
  \bibinfo{author}{\bibfnamefont{K.}~\bibnamefont{Nenkov}},
  \bibinfo{author}{\bibfnamefont{S.}~\bibnamefont{Wurmehl}},
  \bibinfo{author}{\bibfnamefont{A.}~\bibnamefont{Wolter}},
  \bibnamefont{et~al.}, \bibinfo{journal}{Phys.\ Rev.\ B}
  \textbf{\bibinfo{volume}{89}}, \bibinfo{pages}{060504(R)}
  (\bibinfo{year}{2014}).

\bibitem[{\citenamefont{Lee et~al.}(2009)\citenamefont{Lee, Zhang, and
  Wu}}]{Wei-Cheng2009}
\bibinfo{author}{\bibfnamefont{W.-C.} \bibnamefont{Lee}},
  \bibinfo{author}{\bibfnamefont{S.-C.} \bibnamefont{Zhang}}, \bibnamefont{and}
  \bibinfo{author}{\bibfnamefont{C.}~\bibnamefont{Wu}},
  \bibinfo{journal}{Phys.\ Rev.\ Lett.} \textbf{\bibinfo{volume}{102}},
  \bibinfo{pages}{217002} (\bibinfo{year}{2009}).

\bibitem[{\citenamefont{Stanev and Koshelev}(2014)}]{Stanev2014}
\bibinfo{author}{\bibfnamefont{V.}~\bibnamefont{Stanev}} \bibnamefont{and}
  \bibinfo{author}{\bibfnamefont{A.}~\bibnamefont{Koshelev}},
  \bibinfo{journal}{Phys.\ Rev.\ B} \textbf{\bibinfo{volume}{89}},
  \bibinfo{pages}{100505} (\bibinfo{year}{2014}).

\bibitem[{\citenamefont{Hu and Wang}(2012)}]{Xiao2012}
\bibinfo{author}{\bibfnamefont{X.}~\bibnamefont{Hu}} \bibnamefont{and}
  \bibinfo{author}{\bibfnamefont{Z.}~\bibnamefont{Wang}},
  \bibinfo{journal}{Phys.\ Rev.\ B} \textbf{\bibinfo{volume}{85}},
  \bibinfo{pages}{064516} (\bibinfo{year}{2012}).

\bibitem[{\citenamefont{Dias and Marques}(2011)}]{Dias2011}
\bibinfo{author}{\bibfnamefont{R.}~\bibnamefont{Dias}} \bibnamefont{and}
  \bibinfo{author}{\bibfnamefont{A.}~\bibnamefont{Marques}},
  \bibinfo{journal}{Supercond.\ Sci.\ Technol.} \textbf{\bibinfo{volume}{24}},
  \bibinfo{pages}{085009} (\bibinfo{year}{2011}).

\bibitem[{\citenamefont{Wilson and Das}(2013)}]{Wilson2013}
\bibinfo{author}{\bibfnamefont{B.}~\bibnamefont{Wilson}} \bibnamefont{and}
  \bibinfo{author}{\bibfnamefont{M.}~\bibnamefont{Das}}, \bibinfo{journal}{J.\
  Phys.: Condens.\ Matter} \textbf{\bibinfo{volume}{25}},
  \bibinfo{pages}{425702} (\bibinfo{year}{2013}).

\bibitem[{\citenamefont{Tanaka et~al.}(2013)\citenamefont{Tanaka, Yanagisawa,
  and Nishio}}]{Tanaka2013}
\bibinfo{author}{\bibfnamefont{Y.}~\bibnamefont{Tanaka}},
  \bibinfo{author}{\bibfnamefont{T.}~\bibnamefont{Yanagisawa}},
  \bibnamefont{and} \bibinfo{author}{\bibfnamefont{T.}~\bibnamefont{Nishio}},
  \bibinfo{journal}{Physica C: Superconductivity}
  \textbf{\bibinfo{volume}{485}}, \bibinfo{pages}{64 } (\bibinfo{year}{2013}),
  ISSN \bibinfo{issn}{0921-4534},
  \urlprefix\url{http://www.sciencedirect.com/science/article/pii/S092145341200370X}.

\bibitem[{\citenamefont{Stanev and Te\ifmmode \check{s}\else
  \v{s}\fi{}anovi\ifmmode~\acute{c}\else \'{c}\fi{}}(2010)}]{Stanev2010}
\bibinfo{author}{\bibfnamefont{V.}~\bibnamefont{Stanev}} \bibnamefont{and}
  \bibinfo{author}{\bibfnamefont{Z.}~\bibnamefont{Te\ifmmode \check{s}\else
  \v{s}\fi{}anovi\ifmmode~\acute{c}\else \'{c}\fi{}}}, \bibinfo{journal}{Phys.
  Rev. B} \textbf{\bibinfo{volume}{81}}, \bibinfo{pages}{134522}
  (\bibinfo{year}{2010}),
  \urlprefix\url{https://link.aps.org/doi/10.1103/PhysRevB.81.134522}.

\bibitem[{\citenamefont{Yerin et~al.}(2013)\citenamefont{Yerin, Drechsler, and
  Fuchs}}]{Yerin2013}
\bibinfo{author}{\bibfnamefont{Y.}~\bibnamefont{Yerin}},
  \bibinfo{author}{\bibfnamefont{S.-L.} \bibnamefont{Drechsler}},
  \bibnamefont{and} \bibinfo{author}{\bibfnamefont{G.}~\bibnamefont{Fuchs}},
  \bibinfo{journal}{Journal of Low Temperature Physics}
  \textbf{\bibinfo{volume}{173}}, \bibinfo{pages}{247} (\bibinfo{year}{2013}),
  ISSN \bibinfo{issn}{1573-7357},
  \urlprefix\url{http://dx.doi.org/10.1007/s10909-013-0903-9}.

\bibitem[{\citenamefont{Zhitomirsky and Dao}(2004)}]{Zhitomirsky2004}
\bibinfo{author}{\bibfnamefont{M.}~\bibnamefont{Zhitomirsky}} \bibnamefont{and}
  \bibinfo{author}{\bibfnamefont{V.}~\bibnamefont{Dao}},
  \bibinfo{journal}{Phys.\ Rev.\ B} \textbf{\bibinfo{volume}{69}},
  \bibinfo{pages}{054507} (\bibinfo{year}{2004}).

\bibitem[{\citenamefont{Moor et~al.}(2013)\citenamefont{Moor, Volkov, and
  Efetov}}]{Moor2013}
\bibinfo{author}{\bibfnamefont{A.}~\bibnamefont{Moor}},
  \bibinfo{author}{\bibfnamefont{A.}~\bibnamefont{Volkov}}, \bibnamefont{and}
  \bibinfo{author}{\bibfnamefont{K.}~\bibnamefont{Efetov}},
  \bibinfo{journal}{Phys.\ Rev.\ B} \textbf{\bibinfo{volume}{88}},
  \bibinfo{pages}{224513} (\bibinfo{year}{2013}).

\bibitem[{rem({\natexlab{a}})}]{remark2}
\bibinfo{note}{For two-band superconductors it was proven
  \cite{Kuplevakhsky2011} that nontrivial topological states, which arise
  exactly for $n_1 \neq n_2$ are thermodynamically metastable and don't
  correspond to the ground state of the superconducting system. Hence, it is
  reasonable to apply such a type of condition also to the case of three-band
  superconductivity considered here and put $n_1=n_2=n_3=n$. Since the subject
  of our consideration is solely a homogeneous current state, here we don't
  consider the presence of topological defects like phase solitons. In fact,
  solitonic solutions can be studied also within our approach. But as yet as
  the question about of their local or global thermodynamic stability is still
  under debate (unfortunately, the arguments provided in Ref.\
  \onlinecite{Lin2012} are mostly heuristic in nature and related to system
  with open boundary conditions). A consideration of such topological defects
  in the case of periodic boundary conditions is out of scope of the present
  paper.}

\bibitem[{\citenamefont{Hanson}(1994)}]{Hanson1994}
\bibinfo{author}{\bibfnamefont{A.}~\bibnamefont{Hanson}},
  \bibinfo{journal}{Not.\ Amer.\ Math.\ Soc.} \textbf{\bibinfo{volume}{41}},
  \bibinfo{pages}{1156} (\bibinfo{year}{1994}).

\bibitem[{\citenamefont{Candelas et~al.}(1985)\citenamefont{Candelas, Horowitz,
  Strominger, and Witten}}]{Candelas1985}
\bibinfo{author}{\bibfnamefont{P.}~\bibnamefont{Candelas}},
  \bibinfo{author}{\bibfnamefont{G.}~\bibnamefont{Horowitz}},
  \bibinfo{author}{\bibfnamefont{A.}~\bibnamefont{Strominger}},
  \bibnamefont{and} \bibinfo{author}{\bibfnamefont{E.}~\bibnamefont{Witten}},
  \bibinfo{journal}{Nuclear Physics B} \textbf{\bibinfo{volume}{258}},
  \bibinfo{pages}{46} (\bibinfo{year}{1985}).

\bibitem[{\citenamefont{Tanaka and Yanagisawa}(2010)}]{tanaka2010b}
\bibinfo{author}{\bibfnamefont{Y.}~\bibnamefont{Tanaka}} \bibnamefont{and}
  \bibinfo{author}{\bibfnamefont{T.}~\bibnamefont{Yanagisawa}},
  \bibinfo{journal}{Journal of the Physical Society of Japan}
  \textbf{\bibinfo{volume}{79}}, \bibinfo{pages}{114706}
  (\bibinfo{year}{2010}).

\bibitem[{\citenamefont{Tanaka}(2010)}]{tanaka2010c}
\bibinfo{author}{\bibfnamefont{Y.}~\bibnamefont{Tanaka}},
  \bibinfo{journal}{Solid State Comm.} \textbf{\bibinfo{volume}{150}},
  \bibinfo{pages}{1980} (\bibinfo{year}{2010}).

\bibitem[{\citenamefont{Huang and Hu}(2015)}]{Huang2015}
\bibinfo{author}{\bibfnamefont{Z.}~\bibnamefont{Huang}} \bibnamefont{and}
  \bibinfo{author}{\bibfnamefont{X.}~\bibnamefont{Hu}},
  \bibinfo{journal}{Phys\. Rev.\ B} \textbf{\bibinfo{volume}{92}},
  \bibinfo{pages}{214516} (\bibinfo{year}{2015}).

\bibitem[{rem({\natexlab{b}})}]{remark}
\bibinfo{note}{We note that the different behavior of the current density is
  not connected solely with the coinciding strength of the interband
  interactions. According to our analysis it holds for different values of the
  interband interaction coefficients, too, except the case where these
  parameters vanish.}

\bibitem[{\citenamefont{Tanaka et~al.}(2011)\citenamefont{Tanaka, Yanagisawa,
  Crisan, Shirage, Iyo, Tokiwa, Nishio, Sundaresan, and Terada}}]{Tanaka2011}
\bibinfo{author}{\bibfnamefont{Y.}~\bibnamefont{Tanaka}},
  \bibinfo{author}{\bibfnamefont{T.}~\bibnamefont{Yanagisawa}},
  \bibinfo{author}{\bibfnamefont{A.}~\bibnamefont{Crisan}},
  \bibinfo{author}{\bibfnamefont{P.}~\bibnamefont{Shirage}},
  \bibinfo{author}{\bibfnamefont{A.}~\bibnamefont{Iyo}},
  \bibinfo{author}{\bibfnamefont{X.}~\bibnamefont{Tokiwa}},
  \bibinfo{author}{\bibfnamefont{T.}~\bibnamefont{Nishio}},
  \bibinfo{author}{\bibfnamefont{A.~.} \bibnamefont{Sundaresan}},
  \bibnamefont{and} \bibinfo{author}{\bibfnamefont{N.}~\bibnamefont{Terada}},
  \bibinfo{journal}{Physica C} \textbf{\bibinfo{volume}{471}},
  \bibinfo{pages}{747} (\bibinfo{year}{2011}).

\bibitem[{\citenamefont{Yanagisawa and Hase}(2013)}]{Yanisagawa2013}
\bibinfo{author}{\bibfnamefont{T.}~\bibnamefont{Yanagisawa}} \bibnamefont{and}
  \bibinfo{author}{\bibfnamefont{I.}~\bibnamefont{Hase}}, \bibinfo{journal}{J.\
  Phys.\ Soc.\ Jpn.} \textbf{\bibinfo{volume}{87}}, \bibinfo{pages}{124704}
  (\bibinfo{year}{2013}).

\bibitem[{\citenamefont{Weston and Babaev}(2013)}]{Weston2013}
\bibinfo{author}{\bibfnamefont{D.}~\bibnamefont{Weston}} \bibnamefont{and}
  \bibinfo{author}{\bibfnamefont{E.}~\bibnamefont{Babaev}},
  \bibinfo{journal}{Phys.\ Rev.\ B} \textbf{\bibinfo{volume}{88}},
  \bibinfo{pages}{214507} (\bibinfo{year}{2013}).

\bibitem[{\citenamefont{Lin and Hu}(2012{\natexlab{b}})}]{Lin2012}
\bibinfo{author}{\bibfnamefont{S.-Z.} \bibnamefont{Lin}} \bibnamefont{and}
  \bibinfo{author}{\bibfnamefont{X.}~\bibnamefont{Hu}}, \bibinfo{journal}{New
  J.\ Phys.} \textbf{\bibinfo{volume}{14}}, \bibinfo{pages}{063021}
  (\bibinfo{year}{2012}{\natexlab{b}}).

\end{thebibliography}

\end{document}